\documentclass{article}[12pt]
\usepackage{jheppub}
\usepackage[utf8]{inputenc}
\usepackage{amsmath}
\usepackage{amssymb}
\usepackage{amsfonts}
\usepackage{amsthm}
\usepackage{physics}
\usepackage{mathtools}
\usepackage[dvipsnames]{xcolor}
\usepackage{xcolor}
\usepackage{rotating}
\usepackage{array}
\usepackage{appendix}
\usepackage{bbold}
\usepackage{tikz}
\usepackage{textcomp}
\usepackage{subcaption}
\usepackage{bbm}

\def\sc{\textit{sc}}

\def\a{\alpha} 

\title{Gaudin Models and Multipoint Conformal Blocks III: Comb channel coordinates and OPE factorisation}

\author{Ilija Buri\'c$^1$,}
\author{Sylvain Lacroix$^2$,} 
\author{Jeremy Mann$^3$,} 
\author{Lorenzo Quintavalle$^3$,} 
\author{Volker Schomerus$^{3,4}$}
\affiliation{$^1$ Department of Physics, University of Pisa, Largo Pontecorvo 3, I-56127 Pisa, Italy,}
\affiliation{$^2$ Institute for Theoretical Studies, ETH Z\"urich, Clausiusstrasse 47, 8092 Z\"urich, Switzerland,}
\affiliation{$^3$ DESY Theory Group, DESY Hamburg, Notkestrasse 85, D-22603 Hamburg,}
\affiliation{$^4$ II. Institut f\"ur Theoretische Physik, Universit\"at Hamburg, Luruper Chaussee 149, D-22761 Hamburg}
\affiliation{Zentrum f\"ur Mathematische Physik, Universit\"at Hamburg, Bundesstrasse 55, D-20146 Hamburg}

\emailAdd{ilija.buric@df.unipi.it}
\emailAdd{sylvain.lacroix@eth-its.ethz.ch}
\emailAdd{jeremy.mann@desy.de}
\emailAdd{lorenzo.quintavalle@desy.de}
\emailAdd{volker.schomerus@desy.de}

\abstract{We continue the exploration of multipoint scalar comb channel blocks for conformal 
field theories in 3D and 4D. The central goal here is to construct novel comb channel cross 
ratios that are well adapted to perform projections onto all intermediate primary fields. 
More concretely, our new set of cross ratios includes three for each intermediate mixed symmetry tensor exchange. These variables are designed such that the 
associated power series expansion coincides with the sum over descendants. The leading 
term of this expansion is argued to factorise into a product of lower point blocks. We
establish this remarkable factorisation property by studying the limiting behaviour of the 
Gaudin Hamiltonians that are used to characterise multipoint conformal blocks. For six 
points we can map the eigenvalue equations for the limiting Gaudin differential 
operators to Casimir equations of spinning four-point blocks.}

\begin{document}

\maketitle

\section{Introduction}

During the last two decades perturbative and non-perturbative aspects of conformal 
field theories have received widespread attention. Correlation functions of local 
operators are among the most important observables in these theories. Correlators 
of local fields are complicated functions in general, in which the dynamical 
content of the theory is meshed together with mere kinematics. Conformal partial 
wave expansions allow to disentangle dynamics and kinematics. This makes them a 
central analytic tool of conformal field theory. Conformal partial waves, or the 
closely related conformal blocks, are entirely determined by conformal symmetry, 
i.e. they do not contain any dynamical information. When correlation functions 
are expanded in a basis of conformal blocks, the coefficients factorise into a 
product of three-point couplings. The latter are a set of numbers without any 
dependence on the insertion points of the fields and they capture all the
dynamical content of the theory. 

In the case of four-point correlation functions, the relevant conformal blocks are by 
now well understood through the work of Dolan and Osborn and others, see in particular
\cite{Dolan:2000ut,Dolan:2003hv,Dolan:2011dv,Costa:2011dw,SimmonsDuffin:2012uy,
Hogervorst:2013sma,Penedones:2015aga,CastedoEcheverri:2015mkz,Echeverri:2016dun,
Costa:2016hju,Isachenkov:2017qgn,Karateev:2017jgd,Erramilli:2019njx,Fortin:2020ncr} 
and many further references therein. For correlation functions of $N$ local fields
with $N >4$, similar powerful results on conformal blocks do not yet exist, though 
there is some significant recent activity in this area, see for example 
\cite{Rosenhaus:2018zqn,Parikh:2019ygo,Fortin:2019dnq,Parikh:2019dvm,Fortin:2019zkm,
Irges:2020lgp,Fortin:2020yjz,Pal:2020dqf,Fortin:2020bfq,Hoback:2020pgj,
Goncalves:2019znr,Buric:2020dyz,Anous:2020vtw,Fortin:2020zxw,Buric:2021ywo,
Poland:2021xjs}. For this reason, we initiated a novel, integrability based 
approach to multipoint conformal blocks in \cite{Buric:2020dyz}. It extends 
an idea that was advanced initially by Dolan and Osborn, namely to characterise 
conformal blocks through a set of differential equations they satisfy. For 
four-point blocks, these differential equations are eigenvalue equations for 
the set of commuting Casimir differential operators that measure the conformal 
weight and spin of the intermediate field. For a higher number $N$ of insertion 
points, the operators that measure the quantum numbers of the $N-3$ intermediate 
fields are still mutually commuting, but they do not suffice to characterise the 
associated blocks. The challenge to complete the Casimir-like operators into a 
full set of commuting differential operator was solved in \cite{Buric:2020dyz,
Buric:2021ywo}. In these papers we explained how to obtain the missing differential 
operators by taking limits of an $N$-site Gaudin integrable
system~\cite{Gaudin_76a,Gaudin_book83,Feigin:1994in}. 

The set of differential equations and blocks depend on the choice of a so-called 
OPE channel. Among all possible channels, the comb channel and the snowflake 
channel of an $N=6$ point function have received the most attention so far. A detailed 
discussion of possible channel topologies and the corresponding Gaudin limits can 
be found in part I of this series \cite{Buric:2021ywo}. In part II we then went on 
to study so-called vertex systems, i.e. three-point blocks for spinning fields. In the 
case of three-point functions of one scalar and up to two mixed symmetry tensors (MSTs) 
in  $d\leq 4$ we were able to identify the unique Gaudin differential operator. This 
covers all the comb channel vertices 
that can appear in $d=3$ and $d=4$ dimensions. In case the two spinning fields in 
the three-point function are symmetric traceless tensors (STTs) we were able to show 
that the three-point vertex operator is included in the set of five differential operators 
that characterise a five-point function. This was achieved by performing appropriate 
OPE limits to reduce the five-point to a three-point function. At the time, the 
extension of this analysis to comb channel $N$-point blocks with vertices including 
MSTs was hampered because we did not have cross-ratios compatible with the 
relevant limits. This is one of the motivations for part III of our series. Our goal 
here is to address the construction of cross ratios. We will manage to construct a 
set of cross ratios for $N$-point functions in the comb channel that is perfectly 
adapted to OPE factorisations into lower point blocks. The explicit formulas are 
developed for $d \leq 4$ dimensions but the ideas are more general and should 
admit an extension to higher $d$. 
\medskip 

\begin{figure}[htp]
    \centering
    \includegraphics{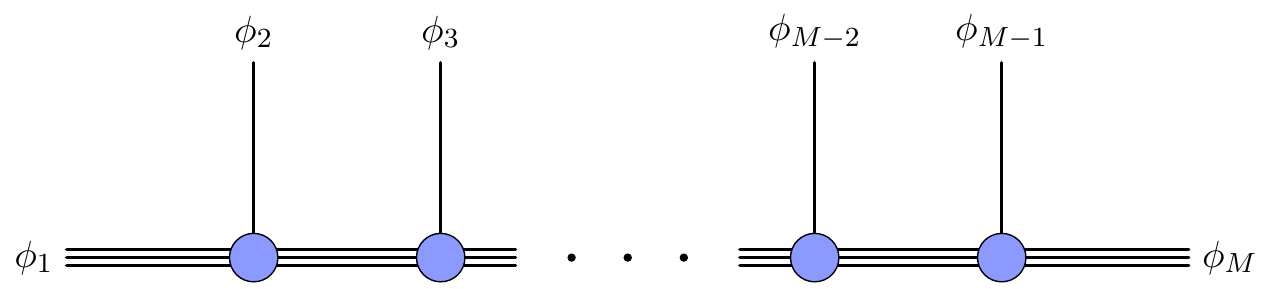}
    \caption{Schematic representation of an $M$-point comb channel OPE diagram in 
    $d=4$. All the external legs at the interior of the comb are scalars, while we 
    allow fields $\phi_1$ and $\phi_M$ to sit in a generic representation.}
    \label{fig:M_point_comb}
\end{figure}

Let us now describe the main new results of this work in some detail. To set up 
some notation we consider the comb channel for $M$ fields in $d=4$, see 
Figure~\ref{fig:M_point_comb}.\footnote{The following discussion is later 
applied to subdiagrams 
of an $N$-point comb channel OPE diagram which is why we do not set $M = N$ 
and will also allow for two of the external fields to carry spin.} In general, 
we can insert arbitrary spinning fields at the external legs, but we shall 
assume that the fields $\phi_j$ on the external legs $j=2, \dots, M-1$ in 
the interior of the comb are scalar fields of conformal weight $\eth_j$. 
The two fields $\phi_1$ and $\phi_{M}$ at the two sides of the comb are 
allowed to carry any spin, i.e. they can be symmetric traceless tensors 
(STTs) or even mixed symmetry tensors (MSTs). We denote the quantum numbers 
of these fields by $\varphi_1 = [\Delta_L,l_L,\ell_L]$ and $\varphi_M = 
[\Delta_R, l_R,\ell_R]$. Here the subscripts $L$ and $R$ stand for `left' \
and `right', respectively, corresponding to the position in the OPE diagram. 
Note that STTs correspond to fields with $\ell = 0$ 
and scalar fields are obtained if we also set $l=0$. The intermediate fields 
that appear along the horisontal lines of the comb are labeled by $[s]$ with 
$s = 1, \dots, M-3$. We may think of $[s] = \{s+1,s+2\}$ as a pair of 
consecutive integers that enumerate the two external scalar fields we attach 
on the two sides of the internal link. The associated intermediate fields 
$\Phi_{[s]}$ possess quantum numbers $\varphi_{[s]} = [\Delta_s,l_s,\ell_s]$ 
with non-vanishing $\ell_s$ for generic cases. Only $\phi_1$ being scalar 
enforces $\ell_1 = 0$ at the first internal leg and similarly we have 
$\ell_{M-3}=0$ in case $\phi_M$ is scalar. When $M > 4$ the total number of 
cross ratios for $M$-point functions 
with $M-2$ scalar and two spinning insertions is given by 
\begin{equation} \label{eq:ncr}
n^{M}_\textit{cr} =  4(M-3) + 1 - 2\delta_{l_L=0=\ell_L} 
- 2 \delta_{l_{R}=0=\ell_{R}}\ . 
\end{equation} 
The subtractions correspond to the cases in which either one or both of the 
fields $\phi_1,\phi_{M}$ are scalar. An $M=3$-point function (vertex) with one 
scalar external field has no cross ratios unless the other two fields are both 
spinning in which case there is a unique cross ratio, see \cite{Buric:2021ttm}. 
For $M=4$ points with at least two scalar insertions one has 
\begin{equation} 
n^{M=4}_\textit{cr} =  5 - 2\delta_{l_L=0=\ell_L} - 2 \delta_{l_{R}=0=\ell_{R}}
+ \delta_{l_L=0=\ell_L} \delta_{l_{R}=0=\ell_{R}}\ . 
\end{equation} 
Note that the application to a four-point function of four scalar fields gives 
$n^{M=4}_\textit{cr} = 5 - 4 + 1 = 2$, i.e. there are two cross ratios in this 
case as is well known. 

The comb channel Hamiltonians are relatively easy to construct, at least in 
principle. In order to do so, we employ the first order differential operators 
$\mathcal{T}_{j,\a}, j=1, \dots M,$ that describe the behaviour of a 
primary field $\phi_j (x_j)$ under the conformal transformation generated 
by the generators $T_\a$ of the conformal algebra. In addition, let us 
also define 
\begin{equation} 
\mathcal{T}_{[s],\a} = \sum_{k=1}^{s+1} \mathcal{T}_{k,\a} \ . 
\end{equation} 
The Casimir differential operators $\mathcal{D}_p^{s}, s= 1,\dots, M-3$ 
are obtained if we construct the $p^{th}$-order Casimir element in terms of
the first-order operators $\mathcal{T}_{[s],\a}$. For generic comb
channel links in $d=4$, the integer $p$ assumes the values $p=2,3,4$. In 
case the field $\phi_1$ is a scalar, the first link only carries two 
quantum numbers and hence there must be one relation between the three 
Casimir elements so that one can restrict to $p=2,4$. A similar statement 
holds in case the field $\phi_{M}$ is scalar. In addition, we also have 
fourth order vertex differential operators of the form
\begin{equation}
\mathcal{V}_{s}^4 = \kappa^{\a_1 \dots \a_4}_4 \mathcal{S}_{s,\a_1} 
\cdots \mathcal{S}_{s,\a_4}, \qquad 
\mathcal{S}_{s,\a} = \mathcal{T}_{s+1,\a} - \mathcal{T}_{[s-1],\a} 
\end{equation}
for $s=1, \dots, M-2$. The operators $\mathcal{V}_1^4$ and $\mathcal{V}_{M-2}^4$ 
can be expressed in terms of the Casimir differential operators in case $\phi_1$
and $\phi_{M}$ are both scalar. So the number of differential operators 
we have constructed here coincides with the number $n^{M}_\textit{cr}$ of cross 
ratios. As we have shown in \cite{Buric:2021ywo}, these operators are all 
independent and they are mutually commuting. Let us note that the set of operators 
with these properties is not unique, of course. In our discussion of the 6-point 
function we will work with a set that is slightly different from the one we 
described here. 
\smallskip 

The joint eigenfunctions of these operators depend on the weights $\eth_j, j=2, 
\dots, M-1$ of the external scalar fields as well as the quantum numbers $\varphi_1 
= [\Delta_L,l_L,\ell_L]$ and $\varphi_{M} = [\Delta_{R},l_{R},\ell_{R}]$ of the two 
fields $\phi_1$ and $\phi_{M}$, respectively. Of course, they also depend on the 
eigenvalues of the differential operators. We parameterise these eigenvalues through 
the quantum numbers $\varphi_{[s]} = [\Delta_s,l_s,\ell_s], s = 1, \dots, M-3$ 
of the internal primaries as well as the eigenvalues $\tau_s, s=1, \dots, M-2,$ 
of the vertex differential operators $\mathcal{V}^s_4$. The latter correspond to 
a choice of tensor structures at the vertices. These wave functions are denoted by 
\begin{equation}
\Psi^{\varphi_1,\eth_j,\varphi_{M}}_{[\Delta_s,l_s,\ell_s;\tau_s]} = 
\Psi^{\varphi_1,\eth_j,\varphi_{M}}_{[\Delta_s,l_s,\ell_s;\tau_s]}(u)\  
\end{equation} 
where $u$ denotes any set of $n^M_\textit{cr}$ independent cross ratios. While the 
construction of the $n_\mathit{cr}$ differential equations these functions satisfy 
is fully algorithmic, see previous paragraph, the resulting expressions are rather 
lengthy in general, see e.g. \cite{Buric:2021ywo} for some examples. On the other 
hand, there are a few cases for which one obtains well-known differential operators. 
For $M=3$ with two spinning fields $\phi_1, \phi_3$, the unique vertex differential 
operator was shown in \cite{Buric:2021ttm} to coincide with the lemniscatic elliptic
Calogero-Moser-Sutherland Hamiltonian discovered by Etingof, Felder, Ma and Veselov 
in \cite{etingof2011107}. 
The most well-known system appears for $M=4$ when all the fields $\phi_i$ are scalar. 
In this case the resulting Hamiltonians famously coincide with those of a 2-particle 
hyperbolic Calogero-Sutherland model of type $BC_2$, \cite{Isachenkov:2016gim}. The 
associated eigenvalue equations turn out to be equivalent to the Casimir equations 
for scalar four-point blocks that were calculated and analysed by Dolan and Osborn
\cite{Dolan:2003hv}. The corresponding eigenfunctions have been studied extensively. 
In mathematics this was initiated by the work of Heckman and Opdam \cite{Heckman-Opdam}. 
The most relevant mathematical results were later re-derived independently in physics, 
starting with the work of Dolan and Osborn \cite{Dolan:2000ut,Dolan:2003hv,Dolan:2011dv}.
Continuing with $M=4$ the next step is to include cases in which one or both of the 
fields $\phi_1$ and $\phi_4$ carry spin. Systems of this type have been studied in 
the physics literature by \cite{Costa:2011mg,Costa:2011dw,Echeverri:2016dun}.  In 
particular so-called seed conformal blocks in $d=4$ dimensions have been characterised 
through a set of Casimir differential equations. 
The solution for these special blocks was developed in the same papers and extensions 
to more general blocks in \cite{Karateev:2017jgd}. Alternatively, it is also possible 
to derive Casimir differential equations within the context of harmonic analysis of 
the conformal group \cite{Schomerus:2016epl,Schomerus:2017eny,Buric:2019dfk}. More 
universally, it is also possible to construct spinning 2-particle Calogero-Sutherland 
Hamiltonians for any choice of spin representations of $\phi_1$ and $\phi_4$, using 
Harish-Chandra's radial component map \cite{HarishChandra} as will be discussed in 
\cite{Buricnew}. The radial component map provides Casimir equations for spinning 
four-point blocks with external fields of arbitrary spin and in any dimension, 
thereby generalising vastly the current status in the physics literature. In spite 
of being so general, the resulting expressions for the universal spinning Casimir 
operators turn out to be surprisingly compact. Nevertheless, a universal solution 
theory has not yet been developed. 
\medskip 

After this preparation we are now able to state the main results of this 
work. They concern conformal blocks for correlation functions of $N$ scalar 
fields. Obviously, the explicit form of the differential operators depends 
very much on the coordinates/cross ratios that are being used. Below we 
shall start with one relatively simple choice that consists of $2(N-3)$ 
four-point cross ratios, $N-4$ five-point cross ratios and $N-5$ six-point 
cross ratios. The total number is $4N-15$ which coincides with the number of 
cross ratios of a scalar $N$-point function in $d=4$ when $N > 4$. These 
initial cross ratios are depicted in Figures \ref{fig:fourpointcratios} 
and \ref{fig:fivesixseven_polynomial}. They turn out to be relatively well 
adapted to performing explicit computations. In particular, one can verify
that all the coefficients of the differential operators are polynomials 
in these cross ratios. For this reason we shall refer to them as 
`polynomial' cross ratios. 

The key to this work is contained in subsection~\ref{sect:six-point_CR} where we 
introduce a new set of independent conformal invariants, first for $N=6$ and then 
more generally for any number $N$ of insertions. The $2(N-3)$ four-point cross ratios 
mentioned above give rise to $N-3$ pairs $(z_r,\bar z_r), r=1, \dots, N-3,$ of 
invariants, one for each internal edge. These are direct generalisations of the 
usual invariants $z,\bar z$ that are used to parameterise four-point cross ratios. 
The five-point cross ratios are then employed to build $N-4$ invariants $w_r, r=2, 
\dots, N-3$, one for each non-trivial vertex. The construction of the $w_r$ is 
an immediate extension of the variable $w$ that we introduced in the study of 
five-point blocks in our previous work \cite{Buric:2021ywo} to complement the variables 
$z_1,\bar z_1,z_2,\bar z_2$. But starting with $N =6$, there exists $N-5$ additional 
independent invariants that involve the six-point cross ratios we described above. 
From these we define new conformal invariants $\Upsilon_r, r = 2, \dots, N-4$, 
one for each internal edge in which an MST can propagate. This invariant is first 
constructed for the unique intermediate MST exchange in a six-point comb channel 
diagram for scalar external fields, see eq.\ \eqref{eq:six_point_polynomial}, and 
then extended to higher numbers $N \geq 6$ of insertions at the end of
section~\ref{sect:six-point_CR}. In the same subsection we also provide a nice 
geometrical interpretation for all the new conformal invariants which we shall 
refer to as comb channel OPE coordinates. 

The association of these invariants with specific links and vertices
is much more than mere counting. Consider a link $r \in \{2, \dots, 
N-4\}$ in which an MST propagates. This link comes with a set of 
three invariants $z_r,\bar z_r, \Upsilon_r$. Our central claim concerning 
OPE factorisation of multipoint blocks can now be formulated after 
rewriting the blocks $\Psi$ in terms of the OPE coordinates $\Psi = 
\Psi(z_r,\bar z_r, \Upsilon_r;w_s)$. When these functions are  expanded 
around $z_b=\bar z_b = 0 = \Upsilon_b$\footnote{Note that the three 
limits do not commute. We take the limit $\bar z_b \rightarrow 0$ first 
before taking $z_b$ and $\Upsilon_b$ to zero. For these last two variables
the order of limits does not matter.} for one particular value of $b 
\in \{2,\dots,N-4\}$ the leading term is claimed to be of the form 
\begin{eqnarray} \label{eq:asymptotics}
& & \Psi^{\eth_i}_{[\Delta_r,l_r,\ell_r;\tau_r]}(z_r,\bar z_r, 
\Upsilon_r;w_s)  =  z_b^{\frac12(\Delta_b+l_b+\ell_b)} 
\bar z_b^{\frac12(\Delta_b-l_b-\ell_b)} \Upsilon_b^{\ell_b} 
\times \\[2mm] & & \quad \quad \quad \times  
\left(\Psi^{\eth_{i\leq b+1},[\Delta_b,l_b,\ell_b]}_{[\Delta_r,l_r,\ell_r;
\tau_s]} (z_r,\bar z_r,\Upsilon_r;w_s)^{r < b}_{s<b}
\times 
\Psi^{[\Delta_b,l_b,\ell_b], \eth_{i>b+1}}_{[\Delta_r,l_r,\ell_r;
\tau_s]}
(z_r,\bar z_r,\Upsilon_r;w_s)^{r>b}_{s\geq b} + 
O(z_b,\bar z_b,\Upsilon_b)\right)\ .  \nonumber 
\end{eqnarray} 
In the first line we have displayed the leading exponents in any of the 
three variables. Note that these are determined by the quantum numbers of 
the exchanged intermediate field $\phi_{[b]}$. In case the latter is an STT, 
i.e. iff $\ell_b = 0$, this leading term is familiar from the theory of 
blocks for four-point functions of scalars. Once this term in the first line 
of the expression is factored out, the remaining function admits a power 
series expansion in the three variables $z_b,\bar z_b$ and $\Upsilon_b$. 
The constant term in this power series expansion turns out to factorise 
into a product of two eigenfunctions of Gaudin Hamiltonians with $M_1 = 
b+2$ and $M_2 = N-b$ sites, respectively. The sub- and superscripts we 
have placed on the eigenfunctions apply to both the dependence of 
quantum numbers and conformal invariants. Let us note that this OPE
factorisation also holds for $b=1$ and $b=N-3$ except that in these 
two cases the quantum number $\ell_b =0$ so that the prefactor in the 
first line only contains powers of $z_b$ and $\bar z_b$. In addition, 
one of the two blocks in the second line is simply a constant. One can 
actually verify such factorisation formulas whenever explicit formulas 
for the blocks are available, e.g.\ for $d=1$ comb-channel blocks which have been constructed in \cite{Rosenhaus:2018zqn}. We have included 
one such explicit check for a six-point function in Appendix~\ref{app:rosenhaus}.  

To prove the remarkable result \eqref{eq:asymptotics} beyond those cases 
in which the blocks are known, the differential operators play a decisive 
role. Strictly speaking, our central claim remains somewhat 
conjectural for $N > 6$. But in the case of $N=6$ we are able to establish 
it rigorously. A scalar six-point function in $d=4$ dimensions depends on
nine cross ratios. We parameterise these through the variables $z_r, \bar 
z_r, r=1,2,3$, $\Upsilon = \Upsilon_2$ and $w_s, s = 1,2$. When we perform 
the limit on the variables $(z_2,\bar z_2,\Upsilon)$ that are associated 
with the internal MST exchange along the central link, the block factorises
into a product of two spinning $M=4$-point blocks with a single spinning 
field and three scalars in each of them. Such spinning four-point blocks 
depend on three variables each. In our special parameterisation these are 
given by $(z_1,\bar z_1;w_1)$ and $(z_3,\bar z_3;w_2)$, respectively. 
As we recalled above, spinning four-point blocks may be characterised as 
solutions of a specific set of differential equations that has been 
worked out at least for some examples in the CFT literature, see in 
particular \cite{Echeverri:2016dun}. As we will announce in this paper, 
the full set of these differential equations can been obtained with the 
help of Harish-Chandra's radial component map \cite{Buricnew}. The strategy 
to prove our factorisation result is to evaluate the limit of the six-site 
Gaudin Hamiltonians as $z_2,\bar z_2,\Upsilon$ are sent to zero 
and to map the resulting operators to the differential operators for 
spinning four-point blocks through an appropriate change of variables. 
Similarly, one can also consider the limit in which the pairs $(z_1,
\bar z_1)$ and $(z_3,\bar z_3)$ are both sent to zero. Our OPE 
factorisation states that the leading term in the resulting expansion 
is given by a spinning four-point block for two scalar and two spinning 
fields. Once again, it is possible to verify this claim by mapping the 
relevant differential operators onto each other. 
\medskip 

Let us now briefly outline the content of each section. The next section is 
entirely devoted to a discussion of cross ratios. After a brief review of the 
two most commonly used sets of cross ratios for four-point functions, we will 
extend both of them to multipoint functions. The usual cross ratios $u,v$ 
can be generalised to higher numbers $N$ of insertion points in such a way 
that the Casimir differential operators for comb channel blocks have 
polynomial coefficients, at least for $N \leq 10$. The relevant 
\textit{polynomial cross ratios} for multipoint functions are defined in 
section~\ref{sect:polynomial_cross_ratios}. While these cross ratios have 
some nice features, they are not 
well adapted to taking OPE limits. For this reason we shall introduce a 
second set of conformal invariants which we dub \textit{OPE cross ratios}. 
We do so for $N=5$ and $N=6$ points first before discussing the case $N > 6$, 
using a nice geometric/group theoretic interpretation of these variables. 
Section~\ref{sect:OPElimits} is devoted to a study of the OPE limits. The 
discussion focuses on $N=6$-point functions. After a brief review of the Gaudin 
Hamiltonians that characterise comb channel blocks, we derive the asymptotic 
behaviour in the first line of eq.\ \eqref{eq:asymptotics} and show that the 
leading term indeed factorises into a product of functions of the respective 
variables. These functions may be characterised through certain differential 
operators which can be obtained by studying the limiting behaviour of the 
original Gaudin Hamiltonians. In particular, it turns out that the Gaudin 
Hamiltonians split into two sets of operators that act on a disjoint subset 
of cross ratios. In section~\ref{sect:CSmodels} we will identify these limiting 
differential operators with the Casimir operators of spinning four-point 
blocks. To this end we briefly sketch the results of \cite{Buricnew} on 
the universal spinning Casimir operators. We cast these results in the 
form of universal spinning Calogero-Sutherland Hamiltonians before we 
compare these operators with those we obtained in section~\ref{sect:OPElimits} 
when taking the OPE limit of multipoint functions. It turns out that the two 
sets of operators coincide. This establishes our result \eqref{eq:asymptotics}, 
including the identification of the leading term with a product of lower point 
conformal blocks. The text concludes with a brief summary along with a list of 
interesting extensions and open problems.

\section{Cross ratios for multipoint correlation functions}

As we have explained in the introduction, there is much freedom in introducing sets of 
independent conformally invariant variables. In this section we introduce two such sets
for multipoint correlation functions. The first one is referred to as polynomial cross 
ratios and it is a direct generalisation of the common four-point cross ratios $u$ and $v$
to scalar correlators with $N > 4$ field insertions. When written in these cross ratios, 
all the $N-3$ quadratic Casimir differential operators that characterise the comb channel
multipoint blocks in sufficiently large dimension $d$ turn out to possess polynomial
coefficients, at least for $N \leq 10$. The second set of conformal invariants 
we introduce in this section is fundamental to all of our subsequent discussion. These new
coordinates are akin to the variables $z$ and $\bar z$ that are widely used for four-point
functions. They possess a large number of remarkable properties. Most importantly for us, 
they behave well under dimensional reductions and when taking OPE limits, which is why we 
shall also refer to them as \textit{OPE cross ratios}. In addition, these variables possess a nice geometric interpretation. 

In the first subsection, the case of $N=4$ will be briefly reviewed to highlight some of 
the properties of the cross ratios $u,v$ and $z,\bar z$ that make them so useful and are 
desirable to maintain as we go to higher number $N$ of insertions. The polynomial cross 
ratios are then introduced in the second subsection. Next, in the third subsection, we 
discuss the OPE coordinates for $N=5$ where there is a single qualitatively new invariant 
that was introduced in \cite{Buric:2021ttm} already. The fourth subsection contains the 
construction of a new invariant that is attached to the central link of the six-point 
comb channel diagram. We introduce this invariant and provide a geometrical interpretation. 
The latter is then used to extend the construction of comb channel invariants in $d=4$ to 
$N > 6$ insertion points.  

\subsection{Prologue: Cross ratios for four-point blocks} 
\label{sect:prologue_CR}

In order to enter the discussion of cross ratios for correlation functions of scalar 
fields, we will begin with the well known case of $N=4$ operators. Famously, there 
exist two cross ratios one can build from their four insertions points $x_i, i=1, 
\dots,4$, 
\begin{equation}
    u=\frac{x^2_{12}x^2_{34}}{x^2_{13}x^2_{24}}\,, \qquad v=\frac{x^2_{14}x^2_{23}}{x^2_{13}x^2_{24}}\,.
\end{equation}
These cross ratios can be represented schematically as in \autoref{fig:fourpointcratios}, 
where we disposed the four points along a square and every colored edge corresponds to a 
scalar product present in the associated cross ratio, with lines that intersect being 
present in the denominator. When written in these four-point cross ratios $u,v$, the second
order Casimir operator takes the form, see eq. (2.10) in \cite{Dolan:2011dv},  
\begin{equation} \label{eq:uvCas}
\mathcal{D}_{(12)}^2 = (1-u-v) \partial_v \left( v \partial_v + a + b\right) + 
u \partial_u\left(2u\partial_u - d\right) - (1+u-v) \left( u \partial_u + v\partial_v 
+ a\right) \left( u \partial_u + v \partial_v + b\right),
\end{equation} 
with the two parameters $2a = \eth_2-\eth_1$ and $2b =\eth_3-\eth_4$ determined by the 
conformal weights $\eth_i$ of the four external scalar fields. We 
observe that in these coordinates, the Casimir operator takes a relatively simple form 
in which all the coefficient functions possess a polynomial dependence on the two cross 
ratios $u$ and $v$. But it also has some less pleasant features. In particular, it is 
not directly amenable to a power series solution in the variables $u,v$. In order to 
formalise this a bit more, let us introduce the notion of a \textit{grade} in some 
variable $w$. We say that a differential operator of the form $c w^n \partial_w^m$ 
has $w$-grade gr$_w(c w^n\partial_w^m) = n-m$. When the grade is applied to some 
linear combination of such simple `monomial' differential operators it returns a 
set of grades, one element for each term. For the grades of the Casimir operator 
\eqref{eq:uvCas} we find 
\begin{equation} 
\textrm{gr}_u \left(\mathcal{D}_{(12)}^2\right) = \{0,1\} \quad , \quad 
\textrm{gr}_v \left(\mathcal{D}_{(12)}^2\right) = \{-1,0,1\}\ . 
\end{equation} 
While the $u$-grade of the individual terms is non-negative, this is not the 
case for the $v$-grade. In other words, when written in the variables $u,v$, 
the quadratic Casimir operator contains simultaneously terms that lower and 
terms that raise the degree of a polynomial in $v$. 

In order to analyse the eigenfunctions of four-point Casimir operators, Dolan and Osborn 
switched to another parameterisation of the cross ratios through the complex variables
$z$ and $\bar z$, 
\begin{equation} 
u = z \bar z \quad , \quad v = (1-z)(1-\bar z) \ . 
\end{equation} 
We point out that the change of variables is not one-to-one since $u$ and $v$ are 
invariant under the action of $\mathbb{Z}_2$ whose non-trivial element exchanges 
$z$ with $\bar z$. Hence, functions of the cross ratios $u$ and $v$ correspond to 
$\mathbb{Z}_2$ invariant functions of $z,\bar{z}$. The invariants $z, \bar{z}$ 
possess a nice geometric interpretation. As is well known, conformal transformations 
can be used to move the insertion points to the special positions $x_2 = 0$, $x_4 = \vec{e}_1$, 
$x_3 = \infty \vec{e}_1$ where $\vec{e}_1$ denotes the unit vector along the first coordinate 
direction of the $d$-dimensional Euclidean space. This choice of a conformal frame is stabilised 
by a subgroup $SO(d-1) \subset SO(d)$ of the rotation group that describes rotations around 
the first coordinate axis. These rotations can be used to move $x_1$ into the plane spanned 
by $\vec{e}_1$ and $\vec{e}_2$. The invariants $z,\bar z$ are the complex coordinates of 
$x_1$ in this plane.  Let us note that in these coordinates it is very easy to implement 
the restriction to $d=1$ for which there exists only one cross ratio, namely $z = \bar z$. 

The geometric interpretation of the $z, \bar z$ coordinates, and in particular 
the simplicity of the reduction to $d=1$, also manifests itself in another property. 
It turns out that the so-called Gram determinant of $N$ insertion points takes a 
particularly simple form when written in $z,\bar z$. Before we state a concrete 
formula, we need to briefly review the concepts of embedding space and Gram 
determinants. The embedding space formalism associates a light-like vector  
$X \in \mathbb{R}^{1,d+1}$ of the form  
$$ X = \left(\frac{1+x^2}{2}, \frac{1-x^2}{2}, x \right) \in \mathbb{R}^{1,d+1}\  
$$
to any point $x \in \mathbb{R}^d$. The associated light rays are in one-to-one 
correspondence with points in $\mathbb{R}$. Note that we have chosen a particular 
representative $X$ of the light ray by fixing the sum of the first two components 
to $X_{-1}+X_{0}=1$.  Given $N$ insertion points $x_i$ we construct $N$ light-like 
vectors $X_i \in \mathbb{R}^{1,d+1}$. These vectors are linearly dependent if 
and only if the associated Gram determinant, i.e.\ the determinant of the matrix 
of scalar products $X_{ij} = \langle X_i,X_j\rangle$, has vanishing determinant. 
For $N=4$ points $x_i \in \mathbb{R}^d$ the associated Gram determinant takes 
the form 
\begin{equation}
    \det(X_{ij})\big|_{4}=(z-\bar{z})^2X_{13}^2X_{24}^2\, . 
\end{equation}
We see that this expression is rather simple when written in terms of the cross
ratios $z,\bar z$, much simpler than its expression in terms of $u,v$. Since any 
four vectors $X_i \in \mathbb{R}^{1,2}$ are linearly dependent, the four-point Gram 
determinant must vanish in $d=1$. This is achieved by setting $z = \bar z$ so that 
all four points lie on a single line, in agreement with our discussion in the 
previous paragraph. The simplicity of the Gram determinant in the $z,\bar z$ 
coordinates means that these are very well suited to implement the dimensional 
reduction. 
\smallskip 

Next we turn to a discussion of the Casimir operator. When the expression we 
spelled out in eq.\ \eqref{eq:uvCas} is rewritten in terms of $z$ and $\bar z$
it acquires the form 
\begin{multline}
    \mathcal{D}_{(12)}^2=2z^2(1-z)\pdv[2]{z}+2\bar{z}^2(1-\bar{z})\pdv[2]
    {\bar{z}}-2(a+b+1)\left(z^2\pdv{z}+\bar{z}^2\pdv{\bar{z}}\right)\\[2mm]
    -2ab(z+\bar{z})+2\varepsilon \frac{z \bar{z}}{z-\bar{z}}\left( (1-z)\pdv{z} 
    - (1-\bar{z})\pdv{\bar{z}}\right)
    \label{eq:zzbCas}
\end{multline}
with $\varepsilon = d-2$. We note that the resulting expression is only slightly 
longer than for the original set $u,v$ of four-point cross ratios. On the other hand, 
its coefficients are no longer polynomial. The main advantage of the $z, \bar z$ 
coordinates is that they admit a rather simple implementation of the OPE limit in 
which we send $\bar{z}\rightarrow 0$ first, followed by the limit $z \rightarrow 0$. 
When $|\bar{z}| < |z|$ we can actually expand the last term in the expression for 
$\mathcal{D}^2_{(12)}$ in a power series. In the resulting expression, all terms 
possess non-negative $\bar{z}$ grade, i.e. 
\begin{equation} 
\textrm{gr}_{\bar z} \left( \mathcal{D}_{(12)}^2\right) \in \{0,1,2,\dots \} \ . 
\end{equation} 
In particular, there is no term in which the derivatives with respect to 
$\partial_{\bar z}$ outnumber the multiplications with $\bar{z}$. Keeping 
only terms of vanishing $\bar{z}$-grade we have 
\begin{equation}
    \mathcal{D}_{(12)}^2 \sim 2z^2(1-z)\pdv[2]{z}+2\bar{z}^2\pdv[2]
    {\bar{z}}-2(a+b+1) z^2\pdv{z} -2abz - 2 \varepsilon \bar{z} \pdv{\bar{z}} + \dots 
    \label{fourpointCasimir12zb0}
\end{equation}
where $\dots$ stand for terms of positive $\bar z$ grade. We can now continue and 
analyse the $z$ grade of the leading term in the previous formula. Obviously, the 
leading terms have non-negative $z$ grade with the terms of vanishing grade given 
by 
\begin{equation}
    \mathcal{D}_{(12)}^2 \sim 2z^2\pdv[2]{z}+2\bar{z}^2\pdv[2]{\bar z}
     - 2 \varepsilon \bar{z} \pdv{\bar{z}} + \dots 
    \label{fourpointCasimir12zzb0}
\end{equation}
where now the $\dots$ contain also terms of positive $z$ grade. Now let us apply 
this discussion to the problem of finding eigenfunctions of the Casimir operator
\begin{equation} 
\mathcal{D}^2_{(12)} \psi_{\Delta,l} (z,\bar z) = 
\left[\Delta(\Delta-d)+l(l+d-2)\right] \psi_{\Delta,l}(z,\bar z) \ . 
\end{equation} 
For the limiting regime in which we replace the Casimir operator by the 
expression in eq.\ \eqref{fourpointCasimir12zzb0} the eigenvalue problem 
is very easy to solve, 
\begin{equation} 
\psi_{\Delta,l}(z,\bar z) \sim z^{\frac{\Delta+l}{2}} \bar z ^{\frac{\Delta-l}{2}} 
c_{\Delta,l} + \dots 
\end{equation} 
where $c_{\Delta,l}$ is a non-vanishing constant factor that is not determined
by the eigenvalue equation and depends on the normalisaton. Since all the terms 
we have omitted from our original Casimir operator have positive grade, we conclude 
that it possesses an eigenfunction of the form 
\begin{equation}
\psi_{\Delta,l}(z,\bar z )\ = \  z^{\frac{\Delta+l}{2}}
\bar z^{\frac{\Delta-l}{2}} F_{\Delta,l} (z,\bar z) = 
z^{\frac{\Delta+l}{2}} \bar z^{\frac{\Delta-l}{2}} \left( 
c_{\Delta,l} + O(z,\bar z)\right) 
\end{equation}
where the function $F$ possesses a power series expansion in $z$ and $\bar z$ with 
non-vanishing constant term $c_{\Delta,l}$.%
\smallskip 

Before we turn to higher number $N >4$ of insertion points we want to summarise 
a few of the desirable properties of the coordinates $z, \bar z$ that seem relevant. 
To begin with, when working with multipoint correlators it is certainly very desirable 
to have simple expressions for the Gram determinant. Note that $N$ points $X_i \in 
\mathbb{R}^{1,d+1}$ are linearly dependent for $N > d+2$. So if we keep the dimension 
$d$ fixed, going to larger values of $N$ will inevitably lead to vanishing Gram 
determinants. Consequently, an $N$-point function in dimension $d < N-2$ lives on 
a subspace within the larger space of cross ratios for $d \geq N-2$. The explicit 
description of this subspace is easiest when we work with coordinates in which 
the Gram determinant factorises into simple functions of the cross ratios. More 
importantly, we would like to find coordinates that are well adapted to the OPE 
limit in the sense we outlined above. For higher multipoint functions this 
means to find coordinates and association of subsets thereof with the 
internal links of the OPE diagram such that eigenfunctions admit a power 
series expansion in all these link variables. For $N > 4$ the leading term 
of these expansions will no longer be constant, of course, but it should 
factorise into a product of functions that are associated with the two 
subdiagrams that are connected by the link. We will indeed be able to 
construct such variables for all $N$-point comb channel diagrams, at least 
in $d=4$ dimensions.  

\subsection{Polynomial cross ratios for comb channel multipoint blocks}
\label{sect:polynomial_cross_ratios}

In this subsection we address the construction of sets of cross ratios which make 
all coefficients of comb channel differential operators polynomial. Because of this 
property, we dub this set \emph{polynomial cross ratios}. We have seen this 
feature before when we wrote the Casimir operator for four-point functions in 
the coordinates $u,v$, see eq.\ \eqref{eq:uvCas}. In this sense, the polynomial 
cross ratios we are about to construct are natural extensions of the four-point 
cross ratios $u,v$.

\begin{figure}[thb]
\begin{subfigure}{0.34\textwidth}
\centering
\includegraphics[height=3.3cm]{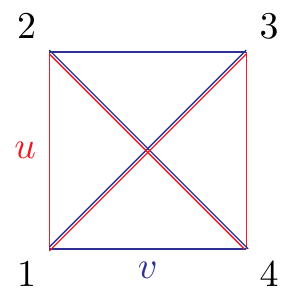}
\caption{\small Four-point cross ratios}
\label{fig:fourpointcratios}
\end{subfigure}
\begin{subfigure}{0.65\textwidth}
\centering
\includegraphics[height=3.3cm]{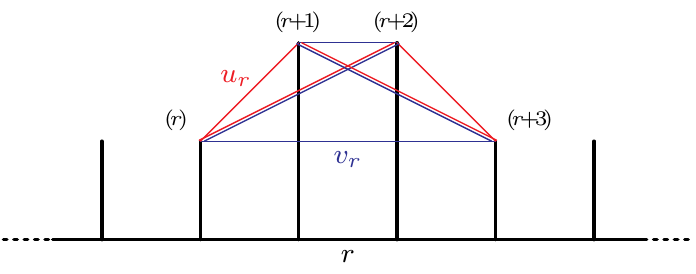}
\caption{\small Construction of four-point cross ratios around internal leg $r$}
\label{fig:four_pt_embedded}
\end{subfigure}
\caption{Schematic representation of the four-point cross ratios $u$ and $v$, where 
intersecting lines correspond to terms in the denominator. The same type of cross 
ratios can be constructed around every internal leg by focusing on the closest four 
points.}
\end{figure}

\begin{figure}[thb]
\begin{subfigure}{.33\textwidth}
\centering
\includegraphics[height=3.3cm]{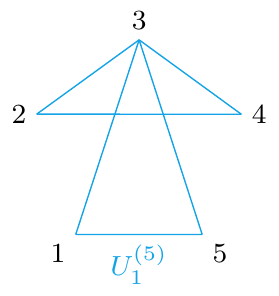}
\caption{\small Five-point cross ratio}
\label{fig:five_pt_polynomial}
\end{subfigure}
\begin{subfigure}{.33\textwidth}
\centering
\includegraphics[height=3.3cm]{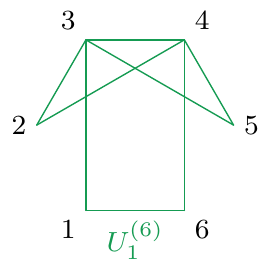}
\caption{\small Six-point cross ratio}
\label{fig:six_pt_polynomial}
\end{subfigure}
\begin{subfigure}{.33\textwidth}
\centering
\includegraphics[height=3.3cm]{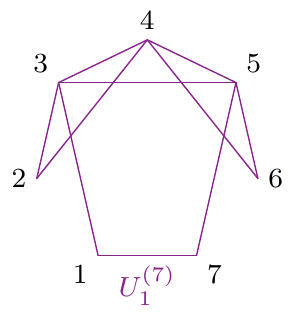}
\caption{\small Seven-point cross ratio}
\label{fig:seven_pt_polynomial}
\end{subfigure}
\caption{Polynomial cross ratios for five, six, and seven point functions. The 
colored lines correspond to scalar products present in the expression of the 
cross ratio, with lines that intersect outside vertices corresponding to terms 
in the denominator.}
\label{fig:fivesixseven_polynomial}
\end{figure}

It is possible to construct the four-point cross ratios of the same type for each 
internal link of the comb channel OPE diagram. Consider the link with label $r= 
1, \dots, N-3$. Then the four nearest neighbor insertion points are $x_i$ with 
$i=r,r+1,r+2,r+3$, see Figure~\ref{fig:four_pt_embedded}. From these we can 
build two four-point cross ratios $u_r,v_r$ using the same expressions as in the 
case of four-point functions, i.e. for an $N$-point comb channel diagram we can 
construct $(N-3)$ sets of $u$, $v$ type cross ratios through
\begin{equation}
    u_r=\frac{X_{r (r+1)}X_{(r+2) (r+3)}}{X_{r (r+2)}X_{(r+1) (r+3)}}\,,\qquad  
    v_r=\frac{X_{r (r+3)}X_{(r+1)(r+2)}}{X_{r(r+2)}X_{(r+1)(r+3)}}\,, \qquad 
    r=1,\dots,N-3\,.
\end{equation}
Here we have used the construction in terms of the embedding space variables
$X_i$, see previous subsection. The $2(N-3)$ cross ratios we have introduced
so far do not suffice to generate all conformal invariants as soon as $d > N-2$. 
We conjecture that a set of cross ratios that makes all coefficients of the 
$N$-point comb channel Casimir operators polynomial in $d \geq N-2$ is obtained 
if we complement the four-point cross ratios $(u_r,v_r)$ introduced above by the 
following set of $m$-point cross ratios 
\begin{equation}
    U^{(m)}_s=\frac{X_{s (s+m-1)}\prod_{j=1}^{m-3}X_{(s+j)(s+j+1)}}
    {\prod_{j=0}^{m-3}X_{(s+j)(s+j+2)}}\,, \qquad s=1,\dots, (N-m+1)\,,
    \qquad m=5,\dots, N.
    \label{mpointpolynomialCR}
\end{equation}
The total number of cross ratios we have introduced is $N(N-3)/2$ which 
coincides with the number of independent cross ratios as long as $d \geq N-2$. 
We checked our claim of polynomial dependence explicitly by verifying that all 
comb channel quadratic Casimir operators that appear for up to $N=10$ external 
scalar fields indeed have polynomial coefficients in the cross ratios we have 
introduced. In addition, we also verified the claim for vertex differential 
operators with $N \leq 6$. We shall often refer to the variables
\eqref{mpointpolynomialCR} as the \textit{$m$-point polynomial cross ratios}, 
since they are constructed around every set of $m$ adjacent points in an 
$N$-point function. The first few examples of these type of cross ratios 
with low values of $m$ are represented schematically 
in Figure~\ref{fig:fivesixseven_polynomial}.

If the dimension $d$ drops below its lower bound or alternatively, if for 
fixed dimension $d$ the number $N$ of insertion points satisfies $N>d+2$, 
there are additional relations between the cross ratios we have introduced. 
These can be found by computing the Gram determinant for the scalar products 
$X_{ij}$. Given $d$, the relations allow to express our $m$-point cross 
ratios $U^{(m)}$ with $m > d+2$ through cross ratios involving a lower 
number of insertion points. In other words, in dimension $d$ the space of 
$N$-point conformal invariants is generated by the cross ratios $U^{(m)}_s$
with $m \leq d+2$. It is easy to verify that the number of such cross ratios
indeed coincides with the expected number $n_\textit{cr}$, see eq.\ 
\eqref{eq:ncr}. 

In $d$ dimensions, there are $N-d-1$ of these $m$-point cross ratios with 
maximal value $m = d+2$. In particular, the first time one of these cross 
ratios is needed is for $(d+2)$-point functions. For example, to 
construct the conformal invariants of an $N$-point function in $d=3$ we
need $m$-point cross ratios with $m=4,5$ only and the first time a 
five-point cross ratio appears is for $N=5$. Similarly, in $d=4$ 
dimensions we work with $m$-point cross ratios for $m=4,5,6$ and all 
these invariants actually appear starting with $N=6$ insertion points. 
Since we are mostly interested in $d=3,4$, it will be sufficient for us
to analyse Casimir operators for correlation functions of $N=5$ and $N=6$ 
scalar fields. 

The set of polynomial cross ratios we have introduced in this subsection leads 
to relatively simple expressions of Casimir operators, but it does not behave 
nicely when taking OPE limits of fields, i.e. the OPE limit cannot simply be 
obtained by taking a limit for a subset of cross ratios to specific values. We 
will now turn to the construction of new variables that are more suitable for 
OPE limits. 
 
\subsection{Five-point OPE cross ratios}
\label{sect:five-point_CR}

We begin our discussion of the new OPE cross ratios with $N=5$. As we reviewed above, 
five insertion points give rise to five independent cross whenever $d \geq 3$. Our 
recipe for the construction of polynomial cross ratios in the previous subsection 
provides us with the following set 
	\begin{equation}
		\begin{gathered}
		u_1=\frac{X_{12} X_{34}}{X_{13} X_{24}}\,, 
		\qquad v_1=\frac{X_{14} X_{23}}{X_{13} X_{24}}\,,\\[2mm]
		u_2=\frac{X_{23} X_{45}}{X_{24} X_{35}}\,, \qquad 
		v_2=\frac{X_{25} X_{34}}{X_{24} X_{35}}\,,
		\end{gathered}
		 \qquad U^{(5)}_1=\frac{X_{15} X_{23}X_{34}}{X_{24} X_{13}X_{35}}\,.
	\label{eq:five_point_polynomial}
	\end{equation}
For this case we did introduce a new parameterisation in~\cite{Buric:2021ywo} 
already through the following set of relations
\begin{equation}
        \begin{gathered}
        u_1=z_1 \bar{z_1}\,,\qquad v_1=(1-z_1)(1-\bar{z}_1)\,,\\[2mm]
    u_2=z_2 \bar{z}_2\,, \qquad v_2=(1-z_2)(1-\bar{z}_2)\,,\\[2mm]
    U^{(5)}_1=w_1(z_1-\bar{z}_1)(z_2-\bar{z}_2)+(1-z_1-z_2)(1-\bar{z}_1-\bar{z}_2)\,.
        \end{gathered}
        \label{eq:frompolynomialtoOPEcoords}
    \end{equation}
Note that the $\mathbb{Z}_2$ symmetry one introduces when passing from $u,v$ to 
$z,\bar z$ in the case of four-point functions is now enhanced to $\mathbb{Z}_2 
\times \mathbb{Z}_2$. In the case of five-point functions the two non-trivial 
generators of this symmetry act by $z_r \leftrightarrow \bar{z}_r, w_1 \rightarrow 
(1-w_1)$ for $r=1,2$. 
When written in the conformal invariant coordinates $z_r,\bar z_r$ and $w=w_1$, the
complexity of the differential operators remains roughly on the same level as for 
the polynomial cross ratios, in the same way as the quadratic Casimir operators 
for $N=4$ which have similar complexity in the two sets of variables, c.f. eqs.\ 
\eqref{eq:uvCas} and \eqref{eq:zzbCas}. But our OPE coordinates for five-point 
functions have a number of additional properties that are worth pointing out. 

\begin{figure}
    \centering
    \includegraphics{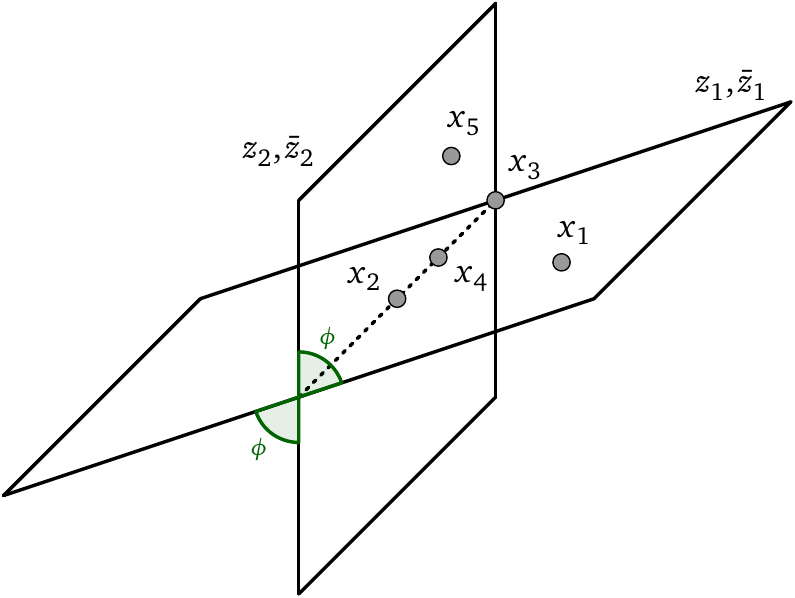}
    \caption{Conformal frame for five points}
    \label{fig:5_points_conformal_plane}
\end{figure}

To begin with, they possess a rather nice geometric interpretation that requires 
going to a certain conformal frame.\footnote{We thank Luke Corcoran for pointing 
out this frame to us.} Using conformal transformations it is possible to move three 
points, let's say $x_2,x_3$ and $x_4$ onto a single line into positions $0,1,
\infty$. Then we can use the remaining rotations transverse to that line in 
order to move $x_1$ into a plane and finally rotations transverse to that 
plane in order to move $x_5$ into some 3-dimensional subspace, i.e. there 
exists a conformal transformation $g^{(5)}$ such that 
\begin{equation}
\begin{gathered}
    g^{(5)}(x_1)=\varrho_1(\cos \theta_1, \sin \theta_1,0, \vec{0})\,, \qquad   
    g^{(5)}(x_2)=(0,0,0,\vec{0})\,,\\[2mm] 
    g^{(5)}(x_3)=(\infty,0,0,\vec{0})\,, \qquad     
    g^{(5)}(x_4)=\vec{e}_1 = (1,0,0,\vec{0})\,,\\[2mm] 
    g^{(5)}(x_5)=e_1-\varrho_2(\cos \theta_2, \sin \theta_2 \cos \phi, 
    \sin\theta_2 \sin \phi, \vec{0})\,. 
    \end{gathered}
    \label{five_point_conformal_frame}
\end{equation}
Here we have parameterised the image point $g^{(5)}(x_1)$ in the plane through an angle 
$\theta_1$ and a distance $\varrho_1$, as usual. Similarly, we have also parameterised the
point $g^{(5)}(x_5)$ in a 3-dimensional space through two angles $\theta_2, \phi$ 
and one distance $\varrho_2$, using $g^{(5)}(x_4)= \vec{e}_1$ as reference point.
In all these expressions, $\vec{0}$ denotes a vector with $d-3$ vanishing components. 
We note that in $d=4$ dimensions, the conformal transformation $g^{(5)}$ is uniquely 
fixed by our choice of frame. It is now easy to compute our new variables $z_r,\bar 
z_r$ and $w_1$ in terms of $\theta_r, \varrho_r$ and $\phi$, 
\begin{equation}
        z_1= \varrho_1 e^{\mathrm{i}\theta_1} \,, \qquad 
        \bar{z}_1=\varrho_1 e^{-\mathrm{i}\theta_1}\,, \qquad    
        z_2= \varrho_2 e^{\mathrm{i}\theta_2} \,, \qquad 
        \bar{z}_2=\varrho_2 e^{-\mathrm{i}\theta_2}\,, \qquad 
        w_1=\sin^2\frac{\phi}{2}\,.
\end{equation}
This is illustrated in Figure~\ref{fig:5_points_conformal_plane}. In particular 
we see that $z_1$, $\bar{z}_1$ and $z_2$, $\bar{z}_2$ describe the two planes 
$x_1x_2x_3x_4$ and $x_2x_3x_4x_5$ respectively, while $w_1$ is directly related 
to the angle $\phi$ between those planes. As we can read off from this picture 
the domain of $w_1$ for Euclidean signature is given by 
\begin{equation}
    w_1\in [0,1]\,.
\end{equation}
The description we provided is valid for $d \geq 3$. As we go down to $d=2$, 
there are no longer enough dimensions in order to have a non-vanishing angle 
$\phi$ between two $2$-planes, i.e. we must set $\phi=0$ or $\phi = \pi$ and 
hence $w_1=0$ or $w_1 = 1$. As in our review of four-point functions, we expect 
to recover these values of $w_1$ as zeroes of the Gram determinant. And indeed, 
the Gram determinant for the five coordinates $X_i$ can be shown to acquire 
the following form 
\begin{equation}
    \det(X_{ij})\big|_{5}=2 \frac{ w_1\left(1-w_1\right) \left(z_1-\bar{z}_1\right)^2 
    \left(z_2-\bar{z}_2\right)^2 X_{13}^2 X_{24}^2 X_{35}^2}{X_{23} X_{34}}\,.
\end{equation}
In addition to the two factors $w_1$ and $w_1-1$ we also notice the zeros 
that appear for $z_r = \bar{z}_r$, i.e. when the four points $x_1x_2x_3x_4$
or $x_2x_3x_3x_5$ lie on a line. 

In section 4.2 of \cite{Buric:2021ttm} we also used these coordinates to 
analyse the OPE limit of our Gaudin differential operators. This analysis 
showed clearly how the variable $w_1$ is naturally associated with the 
degree of freedom that described the choice of tensor structures at the 
internal vertex of a five-point OPE diagram. More specifically, we took the 
OPE limit for the two sets $z_1,\bar z_1$ and $z_2,\bar z_2$ of variables 
that are associated with the two internal links of the OPE diagram. In the 
limit where we take $\bar z_r \rightarrow 0$ first, followed by $z_r 
\rightarrow 0$, the joint eigenfunctions of the five differential 
operators behave as 
\begin{equation}
\psi_{\Delta_r,l_r;\kappa}(z_r,\bar z_r,w_1) \sim 
\prod_{r=1}^2 z_r^{\frac{\Delta_r+l_r}{2}} \bar z_r^{\frac{\Delta_r-l_r}{2}} 
\left(\gamma_{\Delta_r,l_r;\kappa}(w_1) + O(z_r,\bar z_r)\right) \ . 
\end{equation} 
The derivation follows the same steps we outlined in the discussion of 
four-point blocks in the first subsection. But in contrast to the case of 
$N=4$, the leading term $\gamma$ of the power series expansion in $z_r,\bar z_r$ 
is no longer constant but rather an eigenfunction of a single variable vertex 
differential operator for an STT-STT-scalar three-point function which we 
constructed and analysed in \cite{Buric:2021ttm}. The latter was shown to 
arise in the OPE limit of the five-point vertex operator which acts on all 
the five cross ratios before taking the limit. The STT-STT-scalar three-point 
function is determined by conformal symmetry up to a function of a single 
variable. The latter can be constructed in terms of the standard $3$-point 
tensor structures $H$ and $V$. The detailed comparison of the $3$-point with 
the OPE limit of the five-point vertex operators gives 
\begin{equation}
    w_1 \rightarrow 1-\frac{H_{ab}}{V_{a,3b}V_{b,a3}}\,,   
\end{equation}
where $a$ and $b$ are the internal legs on which the OPE limit projects, see 
\cite{Buric:2021ttm} for a detailed discussion. 

\subsection{Six-point OPE cross ratios}
\label{sect:six-point_CR}

After reviewing our parameterisation of five-point cross ratios we now turn to 
a discussion of $N=6$. As long as $d \geq 4$ our set of independent polynomial 
cross ratios consists of 
    \begin{gather}
        u_3=\frac{X_{34} X_{56}}{X_{35} X_{4 6}}\,, \qquad v_3=
        \frac{X_{45} X_{36}}{X_{35} X_{4 6}}\,, \qquad 
        U^{(5)}_2=\frac{X_{26} X_{34}X_{45}}{X_{35}
        X_{24}X_{46}}\,,\label{eq:fourandfiveptpol_insixptf}\\[2mm]
        U^{(6)}_1=\frac{X_{16} X_{23}X_{34}X_{45}}{X_{13}X_{24} X_{35}X_{46}}\,.
        \label{eq:six_point_polynomial}
    \end{gather}
in addition to the five cross ratios already introduced in eq.\ \eqref{eq:five_point_polynomial}. 
While the three cross ratios in the first line are of the same type as those we met in our
discussion of $N=5$, the six-point cross ratio in the second line is fundamentally new. In 
passing to our OPE coordinates it is natural to make use of the map 
\eqref{eq:frompolynomialtoOPEcoords} to transform the cross ratios shared with the 
previously discussed five-point function, while analogously mapping the cross ratios in eq.\  
\eqref{eq:fourandfiveptpol_insixptf} to
\begin{equation}
    \begin{gathered}
    u_3=z_3 \bar{z_3}\,,\qquad v_3=(1-z_3)(1-\bar{z}_3)\,,\\[2mm]
    U^{(5)}_2=w_2(z_2-\bar{z}_2)(z_3-\bar{z}_3)+(1-z_2-z_3)(1-\bar{z}_2-\bar{z}_3)\,. \label{w2def}
    \end{gathered}
\end{equation}
For the six-point variable \eqref{eq:six_point_polynomial}, a new type of mapping is 
necessary. In the same way as the variables $z_r$, $\bar{z}_r$ are associated with 
exchanges of STTs, and the $w_s$ variables are associated 
with specific non-trivial tensor structures sitting at internal vertices of OPE 
diagrams, the new variable we want to introduce should be associated with exchanges 
of Mixed-Symmetry Tensors with two spins, and it should naturally combine with the 
$z_2$, $\bar{z}_2$ cross ratios to make up the three exchanged degrees of freedom 
of the middle link. We propose to introduce this conformal invariant $\Upsilon
= \Upsilon_2$ through the relation
\begin{multline}
    U^{(6)}_1=\Upsilon  \left(z_1-\bar{z}_1\right) \left(z_2-\bar{z}_2\right)
    \left(z_3-\bar{z}_3\right) \sqrt{w_1(1-w_1)w_2(1-w_2)}-w_1 w_2
    \left(z_1-\bar{z}_1\right) \left(\bar{z}_2+z_2\right) \left(z_3-\bar{z}_3\right)\\[2mm]
    +w_1 \left(z_1-\bar{z}_1\right) \left[z_2 \left(1-\bar{z}_3\right)-\bar{z}_2(1-z_3)\right]
    +w_2 \left(z_3-\bar{z}_3\right)\left[z_2 \left(1-\bar{z}_1\right)-\bar{z}_2(1-z_1)\right]\\[2mm]
    +\left[z_2-(1-z_1)(1-z_3)\right] \left[\bar{z}_2-(1-\bar{z}_1)(1-\bar{z}_3)\right]\, .
\label{eq:Upsilondef}
\end{multline}
The new variables $z_r,\bar z_r, w_s$ and $\Upsilon$ admit an action of $\mathbb{Z}_2^3$ 
that leaves the original cross ratios invariant. The nontrivial elements $\sigma_r$ of the 
three $\mathbb{Z}_2$ factors each exchange one of the pairs $z_r \leftrightarrow \bar{z}_r$, 
map $w_s \rightarrow (1-w_s)$ for $r=s,s+1$ and send $\Upsilon$ to $-\Upsilon$.  
\smallskip 

As a first quick test of our proposal, we can compute the six-point Gram determinant. 
When expressed in the OPE coordinates it reads
\begin{equation}
    \det(X_{ij})\big|_{6}=\frac{\left(1-w_1\right) w_1\! \left(1-w_2\right) w_2
    \left(z_1-\bar{z}_1\right)^2 \!\left(z_2-\bar{z}_2\right)^2\! \left(z_3-\bar{z}_3\right)^2\!
    \left(4 z_2 \bar{z}_2-\Upsilon ^2 \left(z_2-\bar{z}_2\right){}^2\right)}{z_2^2 \bar{z}_2^2 }
    \frac{\prod_{i=1}^4X^2_{i,i+1}}{X_{34}^2}.
    \label{gram_6}
\end{equation}
Given the lengthy relation between the six-point cross ratio $U^{(6)}$ and $\Upsilon$
it is very reassuring to see that the Gram determinant now fits into a single line. In 
addition, the new conformal invariant $\Upsilon$ appears in a single factor, combined 
only with the cross ratios $z_2,\bar z_2$. If we reduce the dimension to $d=3$, the 
number of cross ratios drops by one. In our new set of conformal invariants we see 
that $\Upsilon$ can then be expressed in terms of $z_2,\bar z_2$ as 
\begin{equation}
\Upsilon^2 = \frac{4z_2\bar z_2}{(z_2-\bar z_2)^2} \quad    \textit{ for }  \quad 
d = 3. 
\end{equation} 
All these simple relations are quite remarkable. On the other hand they are not yet 
sufficient to fully appreciate our definition of $\Upsilon$. Given what we have 
seen one may for example still wonder why we did not rescale $\Upsilon$ to make 
the last bracket in the Gram determinant equal to $(\Upsilon^2-1)$. While that is 
certainly possible and leads to a nicer geometrical interpretation, the rescaled 
variable would result in more complicated expressions for the asymptotics of 
comb channel blocks in OPE limits, see our discussion in the next section. 

The interpretation of our coordinates proceeds as in the previous subsection. In 
that case, each of the two internal links was associated with a complex plane. We  
used the coordinates $z_1, \bar z_1$ and $z_2,\bar z_2$ to specify two positions 
on these two planes and related the variable $w_1$ to the relative angle between 
the two planes within a 3-dimensional subspace. As we go to $N=6$, the same 
picture applies, but with dimensions raised by one. Instead of the 2-planes in 
3-space, we now have two 3-spaces that are associated with the points $x_1, \dots,
x_5$ and $x_2,\dots,x_6$, respectively. These are embedded in a 4-dimensional 
subspace with the relative angle being measured by a new angle $\varphi$. Each 
of the two 3-spaces contains the configuration of two planes depicted in Figure 
\ref{fig:5_points_conformal_plane}. For the first five points $x_1, \dots, x_5$
this defines the coordinates $\varrho_1,\theta_1,\varrho_2,\theta_2$ and $\phi$ as 
before. We obtain a similar set of coordinates for the second set $x_2, \dots, x_6$. 
Now it is easy to see that one pair of coordinates coincides with the ones from the 
first quintuple of insertion points so that in total we need eight coordinates 
$\varrho_r,\theta_r,\phi_1,\phi_2$ with $r=1,2,3$ to parameterise the configurations 
within each of the 3-spaces. With these coordinates introduced one finds that 
\begin{equation} \label{eq:6ptcrgeom}
z_r := \varrho_r e^{\mathrm{i} \theta_r}, \quad w_{s} := \sin^2 \frac{\phi_s}{2}, 
\quad 
\Upsilon := \pm \mathrm{i} \frac{\cos \varphi}{\sin \theta_2},
\end{equation}
where $r = 1,2,3$ and $s=1,2$. The sign in $\Upsilon$ is conventional, and can 
be absorbed in a shift of the angle $\varphi$. A more formal definition of the 
various geometric parameters on the right hand side will be given in the next 
subsection as part of a more general construction that applies to any number 
$N$ of points in $d=4$ dimensions. 

\subsection{Generalisation to higher number of points} 
 
In order to extend our choice of coordinates to higher number $N$ of insertion points 
in $d=4$ dimensions, it is useful to formalise the construction we have described at 
the end of the previous section. As described in subsection~\ref{sect:five-point_CR}, 
each quintuple of consecutive points $x_s,x_{s+1}, \dots, x_{s+4}$ defines a 
conformal transformation $g^{(5)}_s$ as in eq. \eqref{five_point_conformal_frame}
\begin{align}
&g_s^{(5)}(x_s) =: \varrho_s \vec{n}(\theta_s,0)\ , \quad g_s^{(5)}(x_{s+1}) =: 
(0,0,0,0), 
\nonumber \\[2mm]
&g_s^{(5)}(x_{s+2}) =: (\infty,0,0,0)\ , \quad  \label{eq:confframe5point} 
g_s^{(5)}(x_{s+3}) =: \vec{e}_1 = (1,0,0,0), \\[2mm]
&g_s^{(5)}(x_{s+4}) =: \vec{e}_1-\varrho_{s+1} \vec{n}(\theta_{s+1},\phi_s), \nonumber 
\end{align}
where $s=1,\dots, N-4$ and we defined the unit vectors $\vec{n}$ as  
\begin{equation}
\vec{n}(\theta,\phi):=(\cos \theta, \sin \theta \cos \phi, \sin\theta \sin \phi, 0).
\end{equation}
Thus, to compute $x_6$ in the conformal frame where $g_1^{(5)}(x_1),\dots, 
g_1^{(5)}(x_5)$ are of the form \eqref{five_point_conformal_frame}, we express 
the sixth point as
\begin{equation}
g_1^{(5)}(x_6) = g_1^{(5)}\circ g_2^{(5)-1} (\vec{e}_1 - \varrho_3\,  \vec{n}(\theta_3,\phi_2)) 
\equiv h_{12}^{(5)} (\vec{e}_1 - \varrho_3\,  \vec{n}(\theta_3,\phi_2)).
\end{equation}
By construction, $h_{12}^{(5)}$ is a conformal group element parameterised by the 
cross ratios of the six-point function. In appendix \ref{app:conformal_frames}, we 
compute this conformal transformation and find
\begin{equation}
h_{12}^{(5)-1} = \varrho_2^{-D} \mathcal{I} \sigma_1 e^{-\varphi M_{34}} e^{-\theta_2 M_{12}} 
e^{-\phi_1 M_{23}} e^{P_1},
\label{h_15_to_25}
\end{equation}
where $\mathcal{I}$ is conformal inversion and $\sigma_1:(x^1,x^2,x^3,x^4)\mapsto 
(-x^1,x^2,x^3,x^4)$ is a reflection along the hyperplane orthogonal to the first 
coordinate direction. The explicit action of the element \eqref{h_15_to_25} on 
spacetime points $x$ is given by 
\begin{equation}
h_{12}^{(5)-1}(x) = \varrho_2 \sigma_1 e^{-\varphi M_{34}} e^{-\theta_2 M_{12}} 
e^{-\phi_1 M_{23}} \frac{x-\vec{e}_1}{(x-\vec{e}_1)^2} .
\end{equation}
In particular we read off that the angle $\varphi$ described the relative angle 
between two 3-spaces. It is obvious how to continue these constructions beyond 
$N=6$ points in $d=4$. We continue to introduce comb channel cross ratios $z_r, 
\bar z_r$ and $w_s$ in terms of the polynomial cross ratios through relations 
\eqref{eq:frompolynomialtoOPEcoords} with indices running over $r=1, \dots, N-3$
and $s = 1, \dots N-4$, respectively. Similarly, we introduce $\Upsilon_r$ with 
$r=2, \dots N-4$ through relations of the form \eqref{eq:Upsilondef}. After 
extending our relations \eqref{eq:6ptcrgeom} to a higher number of comb channel 
OPE coordinates we introduce the geometric coordinates as 
\begin{equation} \label{eq:OPEcrgeom}
z_r := \varrho_r e^{\mathrm{i} \theta_r}, \quad w_{s} := \sin^2 \frac{\phi_s}{2}, 
\quad 
\Upsilon_r := \pm \mathrm{i} \frac{\cos \varphi_r}{\sin \theta_{r+1}},
\end{equation}
and define in direct analogy to eq.\ \eqref{h_15_to_25} the conformal transformations 
\begin{equation}
h_{s(s+1)}^{(5)} := g_s^{(5)}\circ g_{s+1}^{(5)-1} = \varrho_{s+1}^{-D} \,\mathcal{I} 
\sigma_1 e^{-\varphi_s M_{34}} e^{-\theta_{s+1} M_{12}} e^{-\phi_s M_{23}}
e^{-\varphi_{s-1} M_{34}} e^{P_1},
\end{equation}
for $s=1,\dots,N-4$. We can thus supplement eqs.\ \eqref{eq:confframe5point} by the 
relations 
\begin{align*}
&g_1^{(5)}(x_6) = h_{12}^{(5)}(\vec{e}_1-\varrho_3\vec{n}(\theta_3,\phi_2)) \ , \quad \quad 
g_1^{(5)}(x_7) = h_{23}^{(5)}\circ h_{12}^{(5)}(\vec{e}_1-\varrho_4 \vec{n}(\theta_4,\phi_3)) \\
& \dots \\
& g_1^{(5)}(x_N) = h_{(N-5)(N-4)}^{(5)}\circ  h_{(N-6)(N-5)}^{(5)}\circ \dots 
\circ h_{23}^{(5)} \circ h_{12}^{(5)}(\vec{e}_1-\varrho_{N-3} \vec{n}(\theta_{N-3},\phi_{N-4})).
\end{align*}
These formulas allow us to compute the location of the insertion points in the conformal frame 
defined by the first five points $x_1, \dots, x_5$, see eq.\ \eqref{eq:confframe5point}, in 
terms of the geometric parameters $\varrho_r,\theta_r,\phi_s$ and $\varphi_r$. The latter 
possess a very simple relation with the OPE cross ratios that we spelled out in eq.\ 
\eqref{eq:OPEcrgeom}. 

\section{OPE limits and factorisation for six-point blocks}
\label{sect:OPElimits}

\begin{figure}[thb]
    \centering
    \includegraphics{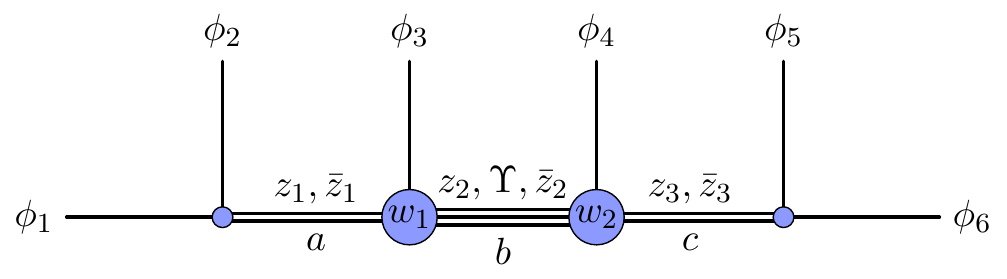}
    \caption{Six point function with external scalars in the comb channel. 
    The $z_i$, $\bar{z}_i$, $w_i$ and $\Upsilon$ type of cross ratios are 
    naturally associated with one particular internal leg or vertex of the 
    OPE diagram.}
    \label{fig:six_point_comb}
\end{figure}

In the previous section we have introduced new conformally invariant coordinates for 
multipoint blocks in $d=4$ dimensions that were naturally attached to the links and 
vertices of a comb channel OPE diagram, see e.g.\  Figure~\ref{fig:six_point_comb} 
for the example $N=6$. To support our choice we provided a nice geometric interpretation 
and, closely related, showed that the Gram determinant for $N=6$ points admits a simple 
factorised expression, see eq.\ \eqref{gram_6}. Recall that the six-point function in 
$d\leq 4$ is the first correlator for which the new link variable $\Upsilon$ appears. 
This makes $N=6$ the decisive case when it comes to testing our cross ratios for 
comb channel blocks in $d=4$. The next two sections are devoted to the most 
important test. 

As we have reviewed in subsection~\ref{sect:prologue_CR}, what makes the cross ratios $z,\bar z$ for 
$4$-point function so useful is the fact that they provide power series expansions 
in the OPE limit where $z,\bar z$ go to zero. One can deduce this important feature 
from the expressions of the Casimir differential operators. Here we want to extend 
this type of analysis to the OPE limits of six-point functions and in particular to 
the limit in which the coordinates $z_2,\bar z_2$ and $\Upsilon$ attached to 
the central link of the comb channel diagram are sent to zero. Our goal is to show 
that in this limit the six-point comb channel blocks possess a power series expansion 
and that the leading term of this expansion factorises into a product of two functions, 
one depending on $z_1,\bar z_1, w_1$, the other on $z_3,\bar z_3, w_2$. 

In our approach we characterise multipoint blocks as eigenfunctions of a complete set of 
commuting differential operators. For $N=6$ comb channel blocks, these 
operators are briefly reviewed in the first subsection. Then we show that the 
OPE limit we are interested in does indeed correspond to sending $z_2, \bar z_2$
and $\Upsilon$ to zero. In the final subsection we then perform the OPE limit on 
the differential operators and show that these operators decouple into two 
independent sets associated with the left and right side of the diagram. We 
also provide concrete expressions for the limiting differential operators. 
These will be further analysed in the next section. 
 
\subsection{Preliminaries on comb channel six-point blocks} 
\label{sect:preliminaries_6_pt}
In this subsection we shall specify all of  our conventions concerning six-point blocks 
and the differential operators we use to characterise them. As usual, any six-point 
correlation function of scalar fields can be split into a product of some homogeneous prefactor $\Omega$, which 
depends on the scaling weights $\eth_i$ and insertion points $x_i$ of the external 
scalar fields, and a function $F$ of the nine cross ratios,  
\begin{equation}
		\expval{\phi_1\phi_2\phi_3\phi_4\phi_5\phi_6}=\Omega^{(\eth_i)}_6(X_i)\,
		F^{(\eth_i)}\!\left(u_1,v_1,u_2,v_2,u_3,v_3,U_1^{(5)},U_2^{(5)},U_1^{(6)}\right)\,. 
		\label{eq:six_point_comb_correlator}
	\end{equation}
The prefactor is not unique. Here we shall adopt the following choice:
\begin{equation}
    \Omega_6^{(\eth_i)}(X_i)=\frac{1}{X_{12}^{\frac{\eth_1+\eth_2}{2}}
    X_{34}^{\frac{\eth_3+\eth_4}{2}}X_{56}^{\frac{\eth_5+\eth_6}{2}}}
    \left(\frac{X_{23}}{X_{13}}\right)^{\frac{\eth_1-\eth_2}{2}}\!
    \left(\frac{X_{24}}{X_{23}}\right)^{\frac{\eth_3}{2}}\!
    \left(\frac{X_{35}}{X_{45}}\right)^{\frac{\eth_4}{2}}\!
    \left(\frac{X_{45}}{X_{46}}\right)^{\frac{\eth_6-\eth_5}{2}}\ . 
\end{equation}
The function $F^{(\eth_i)}$ admits a conformal block decomposition of the form 
\begin{equation}
    F^{(\eth_i)}=\sum_{\Xi}\Lambda_\Xi\, \psi^{(\eth_i)}_{\Xi}  \left(u_r,v_r,U_1^{(5)},
    U_2^{(5)},U_1^{(6)}\right)\,, \quad \textit{where} \quad 
    \Xi=\{\Delta_a,l_a,\Delta_b,l_b,\ell_b,\Delta_c,l_c,\tau_L,\tau_R\}\,
    \label{eq:six_pt_conf_block_dec}
\end{equation}
is a complete set of quantum numbers that includes the weights $\Delta_a,\Delta_b,\Delta_c$ 
and spins $l_a,l_b,\ell_b,l_c$ of the internal fields in the comb channel decomposition, as 
well as two quantum numbers $\tau_L$ and $\tau_R$ that label the choice of tensor structure 
at the two central vertices of the diagram in Figure~\ref{fig:six_point_comb}. We have also 
split each summand into a product of OPE coefficients $\Lambda = \Lambda_\Xi$ and a conformal 
block $\psi_\Xi$. From now on we will drop the labels on $\psi$ unless it is not clear from 
the context which ones they are. 

The six-point comb channel conformal blocks in eq.\ \eqref{eq:six_pt_conf_block_dec} are joint 
eigenfunctions of nine differential operators, as was showed in~\cite{Buric:2021ywo}. These 
include three quadratic Casimir operators, which are constructed for each of the three 
internal links of the OPE diagram as
\begin{align}
    \mathcal{D}^2_{(12)}&= (\mathcal{T}_1+\mathcal{T}_2)_{[AB]}(\mathcal{T}_1+\mathcal{T}_2)^{[BA]}
    =\mathcal{D}^2_{(3456)}\,,\\[2mm]
    \mathcal{D}^2_{(123)}&=(\mathcal{T}_1+\mathcal{T}_2+\mathcal{T}_3)_{[AB]}
    (\mathcal{T}_1+\mathcal{T}_2+\mathcal{T}_3)^{[BA]}=\mathcal{D}^2_{(456)}\,,\\[2mm]
    \mathcal{D}^2_{(56)}&=(\mathcal{T}_5+\mathcal{T}_6)_{[AB]}(\mathcal{T}_5+\mathcal{T}_6)^{[BA]}
    =\mathcal{D}^2_{(1234)}\,. 
\end{align}
Here we have adopted the standard convention to label the generators $\mathcal{T}_\a$ of the 
conformal algebra through pairs $AB$ with $A,B = 1, \dots, d+2$ such that $\mathcal{T}_{AB} 
= - \mathcal{T}_{BA}$. The three quadratic Casimir operators are joined by three quartic ones
which take the following form 
\begin{align}
    \mathcal{D}^4_{(12)}&= (\mathcal{T}_1+\mathcal{T}_2)_{[AB]}
    (\mathcal{T}_1+\mathcal{T}_2)^{[BC]}(\mathcal{T}_1+\mathcal{T}_2)_{[CD]}
    (\mathcal{T}_1+\mathcal{T}_2)^{[DA]}=\mathcal{D}^4_{(3456)}\,,\\[2mm]
    \mathcal{D}^4_{(123)}&=(\mathcal{T}_1+\mathcal{T}_2+\mathcal{T}_3)_{[AB]}
    (\mathcal{T}_1+\mathcal{T}_2+\mathcal{T}_3)^{[BC]}
    (\mathcal{T}_1+\mathcal{T}_2+\mathcal{T}_3)_{[CD]}
    (\mathcal{T}_1+\mathcal{T}_2+\mathcal{T}_3)^{[DA]}=\mathcal{D}^4_{(456)}\,,\\[2mm]
    \mathcal{D}^4_{(56)}&=(\mathcal{T}_5+\mathcal{T}_6)_{[AB]}
    (\mathcal{T}_5+\mathcal{T}_6)^{[BC]}(\mathcal{T}_5+\mathcal{T}_6)_{[CD]}
    (\mathcal{T}_5+\mathcal{T}_6)^{[DA]}=\mathcal{D}^4_{(1234)}\,.
\end{align}
In addition, there is one third-order Pfaffian operator that is assigned to the central 
link, 
\begin{equation}
    \mathcal{D}^3_{(123)}=\epsilon^{ABCDEF}(\mathcal{T}_1+\mathcal{T}_2+\mathcal{T}_3)_{[AB]}
    (\mathcal{T}_1+\mathcal{T}_2+\mathcal{T}_3)_{[CD]}
    (\mathcal{T}_1+\mathcal{T}_2+\mathcal{T}_3)_{[EF]}\,.
\end{equation}
To complete the list of differential operators we finally spell out the two fourth 
order vertex operators,
\begin{align}
    \mathcal{D}_{L,(12)3}^{4,3}&= (\mathcal{T}_1+\mathcal{T}_2)_{[AB]}
    (\mathcal{T}_1+\mathcal{T}_2)^{[BC]}
    (\mathcal{T}_1+\mathcal{T}_2)_{[CD]}(\mathcal{T}_3)^{[DA]}\,,\\[2mm]
    \mathcal{D}_{R,(56)4}^{4,3}&= (\mathcal{T}_5+\mathcal{T}_6)_{[AB]}
    (\mathcal{T}_5+\mathcal{T}_6)^{[BC]}
    (\mathcal{T}_5+\mathcal{T}_6)_{[CD]}(\mathcal{T}_4)^{[DA]}\,.
\end{align}
In the following we will mostly focus on the quadratic Casimir operators. It is rather easy 
to compute the expression of these Casimir operators in the polynomial cross ratios with the 
aid of computer algebra software and verify that all their coefficients 
are indeed  polynomial, as we had claimed in the previous section. The resulting expressions 
for Casimir operators are actually the simplest we have been able to find, simpler than for 
any other set of coordinates. On the other hand, the polynomial cross ratios are not well 
adapted to taking OPE limits, as we will argue in section~\ref{sect:six_point_OPElim_CR}. 
Taking the OPE limit will require passing to the new OPE coordinates introduced in the 
previous section. 

\subsection{The OPE limit from embedding space}
\label{sect:OPE_embedding_space}

Our goal now is to motivate why we expect the sum over descendants in the central intermediate 
link to be encoded in a power series expansion in the variables $z_2,\bar z_2,\Upsilon$. 
The idea here is to prepare the intermediate fields through an operator product expansion 
of either the three fields on the left or the three fields on the right of the central 
link. For the left hand side this amounts to making $x_1,x_2$ and $x_3$ collide. 

It is a little more tricky to understand how the OPE limit is performed once we pass to 
the cross ratios. As an example, let us briefly look at the limit in which $x_1$ and $x_2$ 
come together. In the process we expect to go from a six-point function of scalar fields 
to a five-point function with one STT insertion and four scalars. While the former has 
nine cross ratios, the latter has only seven, i.e. we expect that two cross ratios are 
fixed in the OPE limit. On the other hand, if we apply the limit to the nine polynomial 
cross ratios we find 
\begin{equation}
    u_1\rightarrow 0\,, \qquad v_1\rightarrow 1 \,, \qquad U_1^{(5)}\rightarrow v_2\,, 
    \qquad U_1^{(6)} \rightarrow U_2^{(5)}\, . 
\label{eq:OPElimitinpolynomialCR}
\end{equation}
Of course, this simply means that one needs to consider subleading terms in 
the limiting behaviour of the cross ratios to parameterise the seven cross ratios of 
the resulting five-point function, but it still illustrates how subtle OPE limits are 
in the space of cross ratios. 
\medskip 

In order to analyse the triple OPE limits of our new cross ratios it is advantageous 
to work in embedding space. In the next few paragraphs we will review how to take 
double limits into STTs and triple limits into MSTs. When dealing with computations 
in embedding space we will work in a Poincar\'e patch in which the sum $X_{-1}+X_0$ 
of the first two entries is nonzero. We can then associate the following lightlike 
vector $X \in \mathbb{R}^{1,d+1}$ with the usual Minkowski metric to any insertion 
point $x \in \mathbb{R}^d$, 
\begin{equation}
    X= \left(\frac{1+x^2}{2},\frac{1-x^2}{2},x\right),
    \label{Poincarepatch_X}
\end{equation}
where we use rescalings to set $X_{-1}+X_0=1$. This amounts to working with a 
particular representative of the projective lightray defined by $x$. To describe 
STTs we additionally need polarisation vectors $Z$ which take the form  
\begin{equation}
    Z=(x \cdot z, - x\cdot z,z)\,, \qquad 
    z =(\frac{1-\zeta^2}{2},\mathrm{i}\frac{1+\zeta^2}{2},\zeta) \in \mathbb{C}^d
    \label{Poincarepatch_Z}
\end{equation}
Here $\zeta \in \mathbb{C}^{d-2}$ describes the physical degrees of freedom of the 
polarisation. These are first mapped to a vector $z \in \mathbb{C}^d$ that satisfies 
$z^2 = 0$. We think of $z$ as describing a direction in $\mathbb{C}^{d}$ and use 
rescalings to set $z_1+\textrm{i}z_2 \equiv 1$. In the last step we complement $z$ 
into a vector $Z$ with $d+2$ components. As one can easily verify, the vector $Z$ 
satisfies light cone and transversality conditions of the form $Z^2 = 0 = X \cdot Z$. 

When dealing with MST fields, finally, we need a second polarisation $W$. It has 
$d-4$ physical degrees of freedom which we describe through a vector $\omega \in 
\mathbb{C}^{d-2}$ that satisfies $\omega^2 =0$ and is normalised such that 
$\omega_1+ \textrm{i} \omega_2 \equiv 1$. 
\begin{equation}
    W=(x \cdot w , - \,x\cdot w,w)\,, \qquad w =(\zeta \cdot \omega ,  
    -i\zeta\cdot\omega ,\omega) \in \mathbb{C}^{d}   
    \label{Poincarepatch_W}
\end{equation}
By construction, $W$ is lightlike and transversal to both $X$ and $Z$, i.e. 
$W^2 = 0 = X \cdot W = Z \cdot W$. With this notation in place, we can now spell 
out the procedure to follow to take OPE limits in embedding space for the cases 
of interest below.
\medskip 

To discuss the OPE limit of a pair of scalars inserted at $x_1$ and $x_2$ we use 
their embedding space coordinates $X_1$ and $X_2$. Projecting to the STTs that 
are produced by the OPE of those two scalar fields requires to construct the 
embedding space coordinate $X_{\text{STT}}$ and polarisation $Z_{\text{STT}}$ 
of said fields from the coordinates of the two scalars. This can be achieved by 
taking a lightcone limit $X_1\cdot X_2=0$ first. Once the lightcone condition
is satisfied we introduce 
\begin{equation}
    X_{\text{STT}}=\frac12(X_1+X_2)\,, \qquad
    Z_{\text{STT}}=\frac{1}{(X_{2}-X_1)_1+\mathrm{i} (X_{2}-X_1)_2}(X_2-X_1)\,.
\end{equation}
Note that the prefactor in the definition of $Z_\text{STT}$ ensures that the 
polarisation is normalised such that $z_1+\mathrm{i} z_2 = 1$. Thanks to the 
condition $X_1\cdot X_2=0$, the two vectors we have build from $X_1$ and $X_2$ 
satisfy the usual relations for STT variables, namely they are both square to 
zero, $X_{\text{STT}}^2= 0= Z_{\text{STT}}^2$, and they are transverse to each 
other, i.e.\ $X_{\text{STT}}\cdot Z_{\text{STT}}=0$. So far, we have only assumed 
that the two scalar fields are light-like separated so that $X_1 \cdot X_2 = 0$. 
To complete the OPE limit we can now set $X_2=X_1+\epsilon\, Z_{\text{STT}}$ and 
compute the $\epsilon \rightarrow 0$ limit.

In order to address the triple OPE limit, it remains to discuss the operator product 
of an STT with a scalar field. Let us consider an STT with associated coordinates 
$X_1$, $Z_1$ and a scalar at position $X_2$. If we want to project to the exchange 
of an MST$_2$ produced by the OPE of those two fields, we need to be able to construct 
embedding space coordinates $X_{\text{MST}_2}$ and polarisations $Z_{\text{MST}_2}, 
W_{\text{MST}_2}$ for an MST$_2$ field starting from the degrees of freedom of the 
two initial fields. To do so, we will follow a nested procedure with two limits of 
the type described above. As before, we start by first taking the lightcone limit 
$X_1\cdot X_2=0$ and construct the expressions
\begin{equation}
    X_{\text{MST}_2}=\frac12(X_1+X_2)\,, \qquad
    Z'=\frac{1}{(X_{2}-X_1)_1 + \mathrm{i}(X_2-X_1)_2}(X_2-X_1)\ .
\end{equation}
From here, one can take $X_2=X_1+\epsilon\, Z'$ and compute the $\epsilon\rightarrow 0$ 
limit. This leads temporarily to something described by one coordinate $X_{\text{MST}_2}$ 
and two auxiliary vectors of STT type $Z_1$ and $Z'$. To make this set suitable to describe 
an MST$_2$, we need to further reduce the degrees of freedom of the system and construct a 
variable $W$. This can be achieved by taking the lightcone limit $Z'\cdot Z_1=X_2\cdot Z_1
=0$ and constructing
\begin{equation}
    Z_{\text{MST}_2}=\frac12 (Z'+Z_1) \,, \qquad
    W_{\text{MST}_2}=\frac{1}{(Z'-Z_1)_3 + \mathrm{i}(Z'- Z_1)_4}(Z'-Z_1)\,.
    \label{Wmst2}
\end{equation}
These two vectors indeed satisfy the appropriate conditions for variables associated with an 
MST$_2$, and the normalisation is such that it matches the reference one we spelled out in 
the introduction. At this point we can complete the OPE limit by writing $Z'=Z+\epsilon W$ 
and taking $\epsilon \rightarrow 0$.
\medskip 

Let us now come back to the cross ratios and analyse their behaviour when we take the OPE 
limit. This is particularly simple if we take the OPE limit for the two scalar fields $\phi_1$
and $\phi_2$ in which case one find that $\bar z_1$ and $z_1$ both tend to zero while all other 
cross ratios remain finite. A similar statement holds for the OPE limit of the two scalar field 
$\phi_5$ and $\phi_6$. It is less straightforward to understand the leading behaviour for 
exchanges of a certain MST$_2$ for the internal leg in the middle. To study this, let us 
start by taking first the OPE limit on the left of side of the OPE diagram and reducing 
to a five-point function of fields $\mathcal{O}_a$, $\phi_3,\dots,\phi_6$. Here, 
the OPE limit for leg $b$ can simply be cast as a limit for one STT with coordinates $X_a$, 
$Z_a$ and one scalar with coordinate $X_3$, of the form described in section~\ref{sect:OPE_embedding_space}. Following that, it is possible to check that
\begin{equation}
    w_1\stackrel{((12)3)\text{ OPE}}{\scalebox{3.5}[1]{$\longrightarrow$}} \begin{cases}
    1 \qquad \text{if} \qquad (X_a\wedge X_3)\cdot (X_4\wedge X_5)>0\,,\\
    0 \qquad \text{else}\,,
    \end{cases}
\end{equation}
while the cross ratios $z_2$, $\bar{z}_2$ and $\Upsilon$ all tend to zero. On the other hand, 
if we were to take the limit from the right side in the $((65)4)$ order, we would end up with
\begin{equation}
    w_2\stackrel{((65)4)\text{ OPE}}{\scalebox{3.5}[1]{$\longrightarrow$}} \begin{cases}
    1 \qquad \text{if} \qquad (X_2\wedge X_3)\cdot (X_4\wedge X_c)>0\,,\\
    0 \qquad \text{else}\,,
    \end{cases}
\end{equation}
while  once again $z_2,\bar{z}_2$ and $\Upsilon$ vanish in the limit. This instructs us on 
the fact that the relevant regime to study the projection on exchanges of specific operators 
in the leg $b$ of the six-point function is the part in common for both OPE limits we took 
above, namely $(z_2,\bar{z}_2,\Upsilon)\rightarrow 0$. Taking only these three cross ratios 
to zero, while leaving all others finite, corresponds to a regime in which the two triples 
$(x_1,x_2,x_3)$ and $(x_4,x_5,x_6)$ can each be enclosed in a sphere of radius $r$ which 
is parametrically smaller than the distance $R$ between any two points of the two triples. 
In this limiting regime, we need the six remaining cross ratios to parameterise the 
configuration of insertion points in the two small spheres, see appendix 
\ref{app:bOPE_embedding_space} for some more details.

\subsection{OPE limits of six-point blocks}
\label{sect:six_point_OPElim_CR}

Our main goal in this subsection is to analyse the asymptotics of the 
six-points comb channel blocks in the limit where we send $\bar z_2, 
z_2$ and $\Upsilon$ to zero. We will first study the limiting behaviour 
of the Casimir equation for $\mathcal{D}^2_{(123)}$ under the assumption of 
a leading power-law behaviour of the form
\begin{equation}
   \Psi(z_s,\bar z_s, w_r,\Upsilon) \sim  
   \bar{z}_2^{p_1} z^{p_2}_2 \Upsilon^{p_3} \left( \psi(z_1,\bar z_1, 
   z_3,\bar z_3, w_1, w_2) + O(z_2,\bar z_2, \Upsilon) \right)  
    \label{eq:asymptotic_ansatz}
\end{equation}
in the three variables for the middle leg. In a similar way to what happens 
for exchanges of STTs, see our review in subsection~\ref{sect:preliminaries_6_pt}, 
the precise powers
depend on the order in which the limits are taken. Taking the limit $\Upsilon
\rightarrow 0$ first turns out to be inconsistent, as it produces divergences 
in the Casimir equation. Instead we first take the $\bar{z}_2 \rightarrow 0$ 
limit followed by the one in $z_2$, in direct analogy to the $N=4,5$-point 
functions. Alternatively, we could also send $z_2$ to zero first, but given 
the symmetry of the cross ratios under $z \leftrightarrow \bar{z}$ and $w 
\leftrightarrow (1-w)$ this is a mere issue of convention. Once this limit is 
performed, the order in which the remaining two are performed turns out 
to be irrelevant and one finds 
\begin{equation}
    \bar{z}_2^{-p_1} z^{-p_2}_2 \Upsilon^{-p_3} \mathcal{D}_{(123)}^2
  \bar{z}_2^{p_1} z^{p_2}_2 \Upsilon^{p_3} 
   \ \overset{\bar{z}_2 \rightarrow 0}{\underset{z_2,\Upsilon \rightarrow 0}{\longrightarrow}} \ 
-2\left(d\,  p_1-p_1^2-p_2^2+\left(p_3+1\right)
    (p_2- p_1)-p_3\left(p_3-1\right) \right) + \dots 
\label{eq:limitCas123}
\end{equation}
where we indicated the order of limits by placing the first one above the 
arrow and the remaining two below. As before, the $\dots$ correspond to higher 
order terms in $z_2,\bar{z}_2$ and $\Upsilon$. This behaviour, in which the 
leading term of the second order Casimir differential operator for the central 
link is a constant was what we were going for when we introduced the OPE 
coordinates. Now we see that we have indeed achieved a first important goal.  

Of course, we expect the constant term we just computed to match the 
eigenvalue of the quadratic Casimir element in the MST$_2$ representation 
of the exchanged intermediate field. The latter is related to the weight 
and spin labels of said fields as 
\begin{equation}
    C^2(\Delta_b,l_b,\ell_b) = \Delta_b(\Delta_b-d)+l_b(l_b+d-2)+\ell_b(\ell_b+d-4)\, . 
    \label{eq:Cas_b_eigenvalue}
\end{equation}
Equating this with the constant we computed in eq.\ \eqref{eq:limitCas123}
we can only match the coefficients in front of the dimension $d$ provided 
that 
\begin{equation*}
    p_1=\frac{\Delta_b-l_b-\ell_b}{2}\,.
\end{equation*}
It is then natural to set the exponent $p_2$ of the variable $z_2$ to be 
\begin{equation}
    p_2=    \frac{\Delta_b+l_b+\ell_b}{2}\,.
\end{equation}
This also ensures  that for $\ell_b=0$ one recovers the usual leading 
behaviour for intermediate STT exchange, see 
subsection~\ref{sect:preliminaries_6_pt}. Requiring finally a full match 
with the Casimir eigenvalue leaves us with two possible solutions for the 
leading behaviour in $\Upsilon$ 
\begin{equation}
    p_3=\ell_b \qquad \text{or} \qquad p_3=l_b+1\,.
\end{equation}
This freedom, which cannot be eliminated by considering higher Casimir 
differential operators, is associated with the invariance of the Casimir 
elements under the action of Weyl transformations. Let us note that the 
two possible solutions correspond to the two possible behaviours in $(1-v)$ 
for the four-point s-channel OPE that distinguish between Euclidean and 
Minkowski conformal blocks \cite{Raben:2018sjl}
\begin{equation}
    (1-v)^{l}\,, \qquad \text{or} \qquad (1-v)^{1-\Delta}\,,
\end{equation}
modulo an exchange of $-\Delta\leftrightarrow l$ and $l\leftrightarrow \ell$.  
Along with the interpretation of $\Upsilon$ as a degree of freedom associated 
with MST$_2$ fields, the first solution with $p_3 = \ell_b$ seems to be more 
natural. This choice will later be validated when we compare the limiting 
behaviour of the remaining non-trivial Casimir operators to those of spinning 
four-point blocks.

\begin{figure}[ht]
    \centering
    \includegraphics{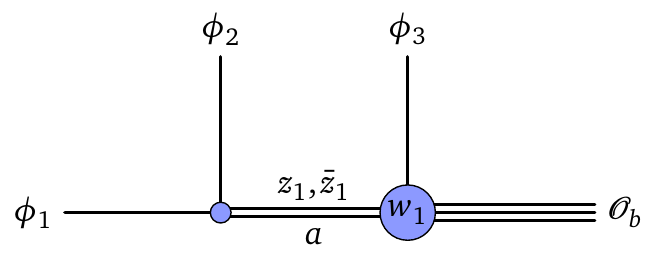}
    \caption{One of the four-point functions obtained in the OPE limit for the middle leg in a six-point function in comb channel. The rightmost field is a Mixed-Symmetry Tensor with two spin indices and the exchanged field is a Symmetric Traceless Tensor.}
    \label{fig:one_MST_four_point}
\end{figure}

Now let us address the second part of our claims. As stated in the introduction 
we want to show that expansion of the conformal block \eqref{eq:asymptotic_ansatz} 
takes the more specific form 
\begin{equation}
    \Psi(z_r,\bar{z}_r,w_1,w_2,\Upsilon)\stackrel{\bar{z}_2,z_2,\Upsilon \rightarrow 0}{\scalebox{1.8}[1.15]{$\sim$}}\bar{z}_2^{\frac{\Delta_b-l_b-\ell_b}{2}}
    z_2^{\frac{\Delta_b+l_b+\ell_b}{2}}\Upsilon^{\ell_b}\,\left(\psi_a(z_1,\bar{z}_1,w_1)
    \psi_c(z_3,\bar{z}_3,w_2)+ \dots \right)\,,
    \label{eq:bOPE_to_4pt}
\end{equation}
in which the leading term splits into a product of two functions of three variables 
each, and to characterise the two factors, one for the left side of the OPE diagram, 
see Figure~\ref{fig:one_MST_four_point}, the other for the right. The proof is a 
nice application of Gaudin integrability, i.e.\ 
our characterisation of multipoint conformal blocks through differential equations. 
Having seen that the differential operators $\mathcal{D}^p_{(123)}, p=2,3,4,$ simply 
acts as multiplication with the value of the associated Casimir elements, we now need 
to study the limiting behaviour of the remaining six differential operators. These 
include two quadratic and two fourth order Casimir operators as well as two vertex 
operators. We will focus our discussion on the quadratic Casimir operators. Very 
remarkably, it turns out that the two quadratic Casimirs $\mathcal{D}^2_{(12)}$ 
and $\mathcal{D}^2_{(56)}$ decouple completely upon taking the OPE limit in the 
central link, 
\begin{equation}
    \mathcal{D}^2_{(12)}\stackrel{b\, \text{OPE}~}{\scalebox{2}[1]{$\longrightarrow$}}\mathcal{D}^2_{a}(z_1,\bar{z}_1,w_1)\,, 
    \qquad \mathcal{D}^2_{(123)}\stackrel{b\, \text{OPE}~}{\scalebox{2}[1]{$\longrightarrow$}}
    C^2(\Delta_b,l_b,\ell_b) \,, \qquad \mathcal{D}^2_{(56)}\stackrel{b\, \text{OPE}~}{\scalebox{2}[1]{$\longrightarrow$}}\mathcal{D}^2_{c}(z_3,\bar{z}_3,w_2)\,.
\end{equation}
Here $b$ OPE denotes the limit in which we take $\bar{z}_2$ to zero followed by $z_2$ 
and $\Upsilon$, as discussed before. Obviously, the expression for $\mathcal{D}^2_{a}$ 
and $\mathcal{D}^2_{c}$ are identical, given the symmetry of the OPE diagram and the 
limiting procedure. Hence it suffices to spell out an expression for $\mathcal{D}^2_a$
which takes the relatively simple form 
\begin{multline} \label{eq:D2MSTlimit}
    \mathcal{D}_a^2=-2 \left(z_1-1\right) z_1^2 \partial_{z_1}^2-2 \left(\bar{z}_1-1\right)
    \bar{z}_1^2 \partial_{\bar{z}_1}^2+\frac{4 \left(w_1-1\right) w_1 z_1 \bar{z}_1 
    \left(w_1 \left(z_1-\bar{z}_1\right)+\bar{z}_1-1\right)}
    {\left(z_1-\bar{z}_1\right){}^2}\partial_{w_1}^2\\[2mm]
    +2 \left(1-w_1\right) w_1 z_1^2 \partial_{z_1}\partial_{w_1}-2 \left(1-w_1\right) 
    w_1 \bar{z}_1^2 \partial_{\bar{z}_1}\partial_{w_1}\\[2mm] 
    +2\left[z_1^2 \left(a+ b-1+\left(w_1-\frac{1}{2}\right)l_b\right)+
    \frac{z_1\bar{z}_1}{z_1-\bar{z}_1}(1-z_1)(d-2)\right]\partial_{z_1}\\[2mm]
    +2\left[\bar{z}_1^2 \left( a+ b-1-\left(w_1 -\frac{1}{2}\right)l_b\right)-
    \frac{z_1\bar{z}_1}{z_1-\bar{z}_1} \left(1-\bar{z}_1\right)(d-2)\right]
    \partial_{\bar{z}_1}\\[2mm]
    \begin{aligned}+2\biggl[ a \left(w_1-1\right) w_1 \left(z_1-\bar{z}_1\right)-&\frac{2 \left(w_1-1\right) w_1 z_1 \bar{z}_1 \left(l_b-1\right)}{z_1-\bar{z}_1}\\[2mm]
    &+\frac{z_1  \bar{z}_1 (d-2) \left(w_1 \left(\bar{z}_1+z_1-2\right)-\bar{z}_1+1\right)}{\left(z_1-\bar{z}_1\right){}^2}\biggr]\partial_{w_1}\end{aligned}\\[2mm] 
    -a \left[\left(2 w_1-1\right) \left(z_1-\bar{z}_1\right) l_b+2 b
    \left(\bar{z}_1+z_1\right)\right]-\frac{z_1 \bar{z}_1  \left(w_1
    \left(z_1-\bar{z}_1\right)+\bar{z}_1-1\right) }{\left(w_1-1\right) 
    w_1 \left(z_1-\bar{z}_1\right){}^2}\ell _b\left(\ell _b+d-4\right)\ . 
\end{multline}
Here the constants $a,b$ are determined by the conformal weights of the external
scalars and $\Delta_b$ through $2a = \eth_2- \eth_1$ and $2b=\eth_3-\Delta_b$.
The expression for $\mathcal{D}_c^2$ looks the same, but with variables $z_3, 
\bar{z}_3,w_2$ instead of $z_1,\bar z_2,w_1$ and parameters $2a = \eth_4-\Delta_b$
and $2b = \eth_3-\eth_4$. 
We have also analysed the fourth order Casimir operators as well as the vertex operators 
and shown that they display the same decoupling. We refrain from spelling out explicit 
expressions here. 

For the time being, all we can do with the explicit expression for $\mathcal{D}^2_{a}$
is to appreciate that the formula looks relatively simple. In the next section we will 
analyse it further and show that it can be mapped to the quadratic Casimir operator 
for a spinning four-point function with three scalar and one MST$_2$ external field. Let 
us note that the blocks for such spinning four-point functions indeed depend on three 
variables, the two $4$-point cross ratios and one additional variable associated with 
the choice of tensor structure at the scalar-STT-MST$_2$ vertex. Notice that our analysis 
implies in particular that conformal partial waves in the limit are polynomials of a 
bounded degree in a variable closely related to $w_1$, given in \eqref{w1-to-XL}, a 
fact that is already non-trivial from the mere definition of OPE cross ratios.
\medskip 

\begin{figure}[ht]
    \centering
    \includegraphics{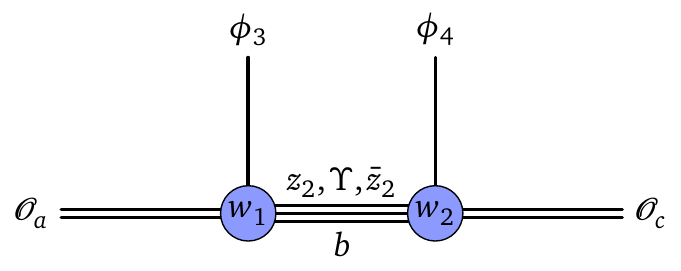}
    \caption{Four-point function obtained from OPE limit on legs $a$ and $c$ of a 
    six-point function in comb channel. Fields at legs $a$ and $c$ are Symmetric 
    Traceless Tensors, while the exchanged field is a Mixed-Symmetry Tensor with 
    two spin indices.}
    \label{fig:two_STT_four_point}
\end{figure}

Before we conclude this section we briefly want to discuss a second OPE limit that 
we have also worked out explicitly. It concerns a setup in which we take 
two OPE limits on the links $a$ and $c$ so that we end up with a four-point function 
of two STT fields and two scalars, see Figure \ref{fig:two_STT_four_point}. 
As explained before, we perform this limit by first sending $\bar{z}_1$ and $\bar{z_2}$
to zero before taking the limits $z_1, z_2 \rightarrow 0$. In the limit, five of the 
nine cross ratios survive and one finds 
\begin{equation}
    \psi(z_1,\bar{z}_1,z_2,\bar{z}_2,z_3,\bar{z}_3,w_1,w_2,\Upsilon)
    \stackrel{\bar{z}_1,z_1,\bar{z}_3,z_3 \rightarrow 0}
    {\scalebox{1.8}[1.15]{$\sim$}}\bar{z}_1^{\frac{\Delta_a-l_a}{2}}
    z_1^{\frac{\Delta_a+l_a}{2}}\bar{z}_3^{\frac{\Delta_c-l_c}{2}}
    z_3^{\frac{\Delta_c+l_c}{2}}\, \left( \psi_b(z_2,\bar{z}_2,w_1,w_2,\Upsilon) + \dots \right)\,.
\end{equation}
The derivation of this limit follows the same steps we carried out in the discussion 
of the OPE limit on the link $b$ above. In particular, one can show that upon taking the 
combined $a$ and $c$ OPE limit, the second order Casimir operators behave as 
\begin{equation}
    \mathcal{D}^2_{(12)},\mathcal{D}^2_{(56)} \stackrel{(a+c)\,
    \text{OPE}~}{\scalebox{3}[1]{$\longrightarrow$}} \text{const}\,, \qquad 
    \mathcal{D}^2_{(123)}\stackrel{(a+c)\, \text{OPE}~}{\scalebox{3}[1]{$\longrightarrow$}}
    \mathcal{D}^2_b(z_2,\bar{z}_2,w_1,w_2,\Upsilon)\,.
\end{equation}
Here `const' denotes the value of the quadratic Casimir element in the STT representations 
of the intermediate fields that are exchanged in the channels $a$ and $c$, respectively. The 
second order Casimir element for the link $b$ reduces to an operator in the five remaining 
variables that can be worked out explicitly, even though the expression is a bit longer 
than in our discussion above. It can be found in the Mathematica notebook we include with 
this publication. In the next section we will show that this 5-variable Casimir operator 
can be mapped to the Casimir operator of a spinning four-point function with two scalars and
two STTs attached on either side of the OPE diagram.

\section{Spinning Calogero-Sutherland models}
\label{sect:CSmodels}

In the last section we have computed the six-point comb channel Casimir differential equations 
in two OPE limits. In the first case, we performed the OPE limit on the central link of the 
OPE diagram and obtained two sets of Casimir operators that act on three cross ratios each. 
The second order Casimir operators were spelled out in eq. \eqref{eq:D2MSTlimit}. The second 
setup involved a combined OPE limit on the left and the right link and it resulted in a set 
of Casimir operators acting on a 5-variable system. In both cases, the resulting limiting  
system is expected to correspond to a spinning four-point correlator, either one involving three 
scalars and one MST or one with two scalars and two STT external fields. The Casimir operators 
for such four-point blocks have been constructed in some examples, see e.g.\ 
\cite{Iliesiu:2015akf,Echeverri:2016dun,Schomerus:2016epl,Schomerus:2017eny}. Here we shall 
report on a very recent observation that all these Casimir operators can be constructed 
from Harish-Chandra's radial component map \cite{Buricnew}. The construction actually 
works for arbitrary spinning four-point functions in any dimension. 

We shall provide a short review of the previous work on the relation between spinning conformal 
blocks and harmonic analysis of spherical functions in the first subsection before we 
spell out the universal spinning Casimir operators in the second. The general construction 
will be worked out explicitly in a number of examples, including the two cases we mentioned 
in the previous paragraph. In the third subsection we then construct an explicit map between 
the OPE limits of Casimir operators obtained in the previous section with the spinning Casimir 
operators to be discussed below, thereby confirming the expectation that e.g. the Casimir 
operator spelled out in eq.\ \eqref{eq:D2MSTlimit} is identical to the Casimir operator 
for a spinning four-point function. 

\subsection{Spherical functions and the radial part of the Laplacian}

As was shown in \cite{Schomerus:2016epl,Schomerus:2017eny,Buric:2019rms}, conformal 
four-point functions admit a realisation as covariant vector-valued functions on the 
conformal group $G \sim SO(d+1,1)$\footnote{We write $G\sim H$ to mean that Lie groups 
$G$ and $H$ are locally isomorphic, i.e. $Lie(G)\cong Lie(H)$.} itself. More precisely, 
there is a bijective 
correspondence between solutions to conformal Ward identities and $K$-spherical 
functions on $G$, where  $K\sim SO(1,1)\times SO(d)$ is the group of rotations 
and dilations. Given two finite-dimensional irreducible representations $\rho_L$ 
and $\rho_R$ of $K$, with carrier spaces $W_{L,R}$, respectively, the space of 
$K$-spherical functions is defined as
\begin{equation}\label{K-spherical-functions}
    \Gamma_{\rho_L,\rho_R} = \{ f : G \xrightarrow{} \text{Hom}(W_R,W_L) 
    \ |\ f(k_L g k_R) = \rho_L(k_L) f(g) \rho_R(k_R)\}, \quad g\in G,\ 
    k_{L,R}\in K\ .
\end{equation}
We may, and occasionally will, identify $\text{Hom}(W_R,W_L)\cong W_L\otimes W_R^\ast$.
Covariance properties of $f$ are then written as
\begin{equation}
    f(k_L g k_R) = (\rho_L(k_L) \otimes \rho_R^\ast(k_R^{-1})) f(g)\ .
\end{equation}
The explicit map between $K$-spherical functions and conformal 
correlators may be found in \cite{Buric:2019dfk,Buric:2020buk}. 

Through the reinterpretation of conformal correlators as $K$-spherical functions on the conformal group, conformal blocks are carried to eigenfunctions of the group Laplacian, thus becoming a subject of harmonic analysis. By definition, spherical harmonics are eigenfunctions of the {\it radial part} of the Laplacian. The latter may be thought of as a differential operator in two variables with matrix coefficients. To explain this, we recall that the conformal group $G$ admits a Cartan decomposition
\begin{equation}
    G = K A_p K\ .
\end{equation}
In other words, almost every\footnote{The set of elements that cannot be factorised has Haar measure zero.} element of $G$ can be factorised as $g = k_L a k_R$, with $k_L,k_R\in K$ and $a\in A_p$. The factor $A_p$ is the two-dimensional abelian group generated by elements $H_1 = \frac12(P_1+K_1)$ and $H_2 = -\frac i2(P_2 - K_2)$, where  $P_\mu, K_\mu$ denote generators or translations 
and special conformal transformations, respectively. Clearly, spherical functions are uniquely determined by their restrictions to $A_p$. Furthermore, such restrictions 
are not arbitrary. Let $M\sim SO(d-2)$ be the centraliser of $A_p$ in $K$. For any $m\in M$, given one decomposition $g = k_L a k_R$, we can form another $g = (k_L m)a(m^{-1}k_R)$ (one can fix the ambiguity by requiring that either $k_L$ or $k_R$ belongs to a particular section of $K/M$). When restricted to $A_p$, spherical functions take values in the space of $M$-invariants inside $\text{Hom}(W_R,W_L)$. Indeed
\begin{equation}\label{m-invariance}
    f(a) = f(m a m^{-1}) = \rho_L(m) f(a) \rho_R(m^{-1})\ .
\end{equation}
Because it commutes with left and right regular representations, the Laplacian $\Delta$ preserves the space $\Gamma_{\rho_L,\rho_R}$. The above comments allow us to regard its restriction to this space as a differential operator that acts on vector-valued functions in two variables.

\subsection{Spinning blocks and Calogero-Sutherland models}

We will now explain how radial parts of the Laplacian are related to spinning Calogero-Sutherland models.

Let $h = \exp (t_i H_i)$ be an element of $A_p$ and write $X'=h^{-1} X h$ for any $X\in\mathfrak{g}$\footnote{We use the notation $\mathfrak{g} = Lie(G)$ and $\mathfrak{g}_c = \mathfrak{g}\otimes\mathbb{C}$ and similarly for Lie algebras of all other groups under consideration.}. The quadratic Casimir of $\mathfrak{g}$ can be written as
\begin{align}
    C_2 & =  H_1^2 + H_2^2 + \coth(t_1+t_2) (H_1 + H_2) + \coth(t_1-t_2)(H_1-H_2) + (d-2)(\coth t_1 H_1 +\coth t_2 H_2)\nonumber \\[2mm]
    & -\frac{D'^2_+ - 2\cosh(t_1+t_2)D'_+ D_+ + D^2_+}{2\sinh^2(t_1+t_2)} - \frac{D'^2_- - 2\cosh(t_1-t_2)D'_- D_- + D^2_-}{2\sinh^2(t_1-t_2)}\label{radial-decomposition-of-Casimir}\\[2mm]
    & + \sum_{a=3}^d \left(\frac{M'^2_{1a}-2\cosh t_1 M'_{1a}M_{1a}+M^2_{1a}}{\sinh^2 t_1} + \frac{M'^2_{2a} -2\cosh t_2 M'_{2a}M_{2a} + M^2_{2a}}{\sinh^2 t_2}\right)- \frac12 M^{ab}M_{ab},\nonumber
\end{align}
where we have introduced $D_\pm = D\pm iM_{12}$ and indices $a,b$ to run over the set $\{3,4,...,d\}$. In the remainder of this section, Latin indices will always be assumed to be in this range. The validity of the last equation can be readily checked and we provide a short derivation in appendix~\ref{app:Eucl_conf_group}. We call the above
equation the radial decomposition of $C_2$. More generally, the radial decomposition of elements in $U(\mathfrak{g}_c)$ may be thought of as an infinitesimal version of the Cartan decomposition. Significance of the radial decomposition of $C_2$ lies in the fact that it allows to directly reduce the Laplacian to any space of $K$-spherical functions. All one needs to do is substitute the generators $H_i$ by partial derivatives $\partial_{t_i}$, the primed generators $x'$, with $x\in \mathfrak{k}_c$, by representation operators $\rho_L(x)$ and the unprimed generators $y\in \mathfrak{k}_c$ by $\rho_R^\ast(-y)$. The fact that the same prescription can be applied for all choices of $\rho_L$ and $\rho_R$ can be captured by defining a universal map
\begin{equation}
    \Pi: U(\mathfrak{g}_c)\to \mathcal{D}(A_p) \otimes(U(\mathfrak{k}_c)\otimes_{U(\mathfrak{m}_c)}U(\mathfrak{k}_c)),
\end{equation}
that assigns to any element of $U(\mathfrak{g}_c)$ a (class of a) differential operator on $A_p$ with coefficients in two copies of $U(\mathfrak{k}_c)$ (here $\mathcal{D}(A_p)$ denotes the algebra of differential operators on $A_p$). The latter is called the Harish-Chandra's radial component map. In practise, for any $u\in U(\mathfrak{g}_c)$, $\Pi(u)$ is computed by radially decomposing $u$\footnote{Almost any element $h\in A_p$ provides an isomorphism of vector spaces $U(\mathfrak{g}_c) \cong U(\mathfrak{a}_{p_c}) \otimes (U(\mathfrak{k}_c) \otimes_{U(\mathfrak{m}_c)} U(\mathfrak{k}_c))$
\begin{equation*}
    \Gamma_h : U(\mathfrak{a}_{p_c}) \otimes U(\mathfrak{k}_c) \otimes U(\mathfrak{k}_c) \xrightarrow{} U(\mathfrak{g}_c),\quad \Gamma_h(H\otimes x\otimes y) = h^{-1} x h\ H\ y\ .
\end{equation*}
The element of $U(\mathfrak{g}_c)$ on the right hand side is said to be in a radially-decomposed form with respect to $h$. Notice that $U(\mathfrak{a}_{p_c})$ is naturally represented by differential operators on $A_p$ with constant coefficients, i.e. $U(\mathfrak{a}_{p_c})\cong\mathbb{C}[\partial_{t_1},\partial_{t_2}]$.} and then replacing elements $x'$, with $x\in U(\mathfrak{k}_c)$, by $x\otimes 1$ and elements $y$, with $y\in \mathfrak{k}_c$, by $1\otimes y$. The replacements here belong to the product of two {\it commuting} copies of $U(\mathfrak{k}_c)$. 

The universal Calogero-Sutherland Hamiltonian is a close cousin of the universal radial part of the Laplacian $\Pi(C_2)$ (see \cite{Stokman:2020bjj,Reshetikhin:2020wep} 
for a recent discussion). The two are related by conjugation, $H = \delta\Pi(C_2)\delta^{-1}$, by the factor
\begin{equation}\label{conjugation-factor}
    \delta(t_i) = \sqrt{\sinh^{d-2} t_1 \sinh^{d-2} t_2 \sinh(t_1+t_2) \sinh(t_1-t_2)}\ .
\end{equation}
This is the essentially unique factor that gives an operator in a Schr\"odinger from, i.e. without first order derivatives in $t_i$. Explicitly, the universal Hamiltonian reads
\begin{align}
     H &= \partial_{t_1}^2 + \partial_{t_2}^2 +\frac{1 - D'^2_+ + 2\cosh(t_1+t_2)D'_+D_+  - D^2_+}{2\sinh^2(t_1+t_2)} + \frac{1 - D'^2_- + 2\cosh(t_1-t_2)D'_-D_- - D^2_-}{2\sinh^2(t_1-t_2)}\nonumber\\[2mm] 
       & + \frac{M'_{1a}M'_{1a}-2\cosh t_1 M'_{1a}M_{1a}+M_{1a}M_{1a}-\frac14(d-2)(d-4)}{\sinh^2 t_1}\label{universal-Hamiltonian}\\[2mm] 
       & + \frac{M'_{2a}M'_{2a}-2\cosh t_2 M'_{2a}M_{2a}+M_{2a}M_{2a} - \frac14 (d-2)(d-4)}{\sinh^2 t_2}- \frac12 M^{ab}M_{ab}- \frac{d^2-2d+2}{2}\ .\nonumber
\end{align}
We have slightly abused the notation here: in eq.\ \eqref{radial-decomposition-of-Casimir} both $M_{1a}$ and $M_{1a}'$ are elements of $U(\mathfrak{g}_c)$ and $M_{1a}' = h^{-1}M_{1a}h$, whereas in eq.\  \eqref{universal-Hamiltonian}, they are elements of $U(\mathfrak{k}_c)\otimes U(\mathfrak{k}_c)$. In accordance with the prescription spelled out above, spinning Calogero-Sutherland Hamiltonians, denoted $H_{\rho_L,\rho_R}$, are obtained from $H$ by substitutions $x'\to \rho_L(x)$ and $y\to\rho_R^\ast(-y)$, with $x,y\in U(\mathfrak{k}_c)$. Equation \eqref{universal-Hamiltonian} is the central one of this section and all applications below emerge from its special cases. Compared to the previous works \cite{Schomerus:2016epl,Schomerus:2017eny}, which computed $H_{\rho_L,\rho_R}$ for some particular representations $\rho_{L,R}$, the new observation of \cite{Buricnew} concerns the universal spin dependence of these Schr\"odinger operators.\footnote{In the most direct way to derive Casimir 
equations for spinning conformal blocks, one would introduce spin by modifying generators of conformal transformations that act on individual fields. Upon reduction to the cross ratio space, this produces additional terms in the Casimir operator. Our procedure circumvents the reduction of spinning degrees of freedom and adds their contribution to the reduced operator directly.}

As a first simple illustration of this universal formula let us briefly discuss the case of scalar functions on $G$. These correspond to four-point functions of scalar fields. For $K$-$K$ invariant functions, i.e. trivial representations $\rho_L = \rho_R = 1$ the Hamiltonian we get reads
\begin{equation*}
    H_0 = \partial_{t_1}^2 + \partial_{t_2}^2 + \frac12\left(\frac{1}{\sinh^2(t_1+t_2)} +\frac{1}{\sinh^2(t_1-t_2)}\right) - \frac{(d-2)(d-4)}{4} \left(\frac{1}{\sinh^2 t_1} +
    \frac{1}{\sinh^2 t_2}\right) - \frac{d^2-2d+2}{2}\ .
\end{equation*}
To make our coordinates on $A_p$ agree with \cite{Schomerus:2016epl,Schomerus:2017eny}, we introduce $u_1 = t_1 + t_2$ and $u_2 = t_1 - t_2$. The Hamiltonian then can be written as 
\begin{equation}
    H_0 = -2\left( H_{PT}^{(0,0)}(u_1) + H_{PT}^{(0,0)}(u_2)+\frac{(d-2)(d-4)}{8}\left(\frac{1}{\sinh^2\frac{u_1+u_2}{2}} +\frac{1}{\sinh^2\frac{u_1-u_2}{2}}\right)\right) - \frac{d^2-2d+2}{2},
\end{equation}
where $H^{(a,b)}_{PT}$ denotes the quantum mechanical P\"oschl-Teller Schr\"odinger operator.\footnote{Conventions for P\"oschl-Teller and Calogero-Sutherland Hamiltonians $H_{PT}^{(a,b)}$, $H_{cs}^{(a,b,\epsilon)}$ agree with \cite{Buric:2021yak}.} The $H_0$ is the Hamiltonian of the hyperbolic $BC_2$ Calogero-Sutherland model with parameters 
$a=0$, $b=0$ and $\epsilon=d-2$, justifying our terminology. More precisely
\begin{equation}
    H_0 = -2 H^{(0,0,d-2)}_{cs} - \frac{d^2-2d+2}{2}\ .
\end{equation}
For non-identical scalar fields, $\rho_L$ and $\rho_R$ are characters of the dilation group $SO(1,1)$ and trivial representations of $SO(d)$. Writing $\rho_L(D)=2a$ and $\rho_R^\ast(D)
=2b$, we obtain the potential
\begin{equation}
    V_{\rho_L,\rho_R} = -\frac{2a^2 - 4\cosh(t_1+t_2)a b + 2b^2}{\sinh^2(t_1+t_2)} - \frac{2a^2 - 4\cosh(t_1-t_2)a b + 2b^2}{\sinh^2(t_1-t_2)}\ .
\end{equation}
The full Hamiltonian is then that of the $BC_2$ Calogero-Sutherland model with parameters $a$, $b$ and $\epsilon=d-2$
\begin{equation}
    H_{\rho_L,\rho_R} = H_0 + V_{\rho_L,\rho_R} = -2 H^{(a,b,d-2)}_{cs} - \frac{d^2-2d+2}{2}\ .
\end{equation}
Wave functions of this operator were constructed in the seminal work of Heckman and Opdam \cite{Heckman-Opdam}. Finally, assume that $\rho_L$ and $\rho_R$ arbitrary finite-dimensional representations. Due to the invariance condition \eqref{m-invariance}, the Hamiltonian 
$H_{\rho_L,\rho_R}$ is restricted to act on functions $F:A_p\to W_L\otimes W_R^\ast$ that satisfy $(\rho_L(M_{ab})+\rho_R^\ast(M_{ab}))F=0$. It is not difficult to see from the expression \eqref{universal-Hamiltonian} that $H_{\rho_L,\rho_R}$ is indeed a well-defined operator on this space. By
writing $\rho_L(x)$ and $\rho^\ast_R(-y)$ as matrices one ends up with a certain matrix Schr\"odinger operator. Here we will follow a slightly different path: representations $\rho_L$ and $\rho_R$ of $\mathfrak{k}_c$ will be written in terms of differential operators that act on finite-dimensional spaces of polynomials. This allows to elegantly impose the $M$-invariance conditions - "spin cross ratios" of previous sections will arise as generators of $M$-invariants in $\text{Hom}(W_R,W_L)$. When prepared in this way, the spinning Hamiltonians may be compared to those obtained through the OPE limit construction. 
\medskip 

We will consider two examples in particular: $1)$ the representation $\rho_L$ is trivial and $\rho_R$ is a mixed symmetry tensor of depth two, and $2)$ both $\rho_L$ and $\rho_R$ are 
symmetric traceless tensors. Here, we are referring to the $SO(d)$ content of these representations. By the dictionary of \cite{Schomerus:2016epl,Buric:2019dfk}, the two cases correspond in conformal field theory to a four-point function of an $\text{MST}_2$ field and three scalars, and two STTs and two scalars, respectively, and hence they are directly relevant for the discussion of the two OPE limits we analysed in the previous section. 

\subsubsection{One \texorpdfstring{$\text{MST}_2$}{MST2} and three scalars}
\label{sect:CS_MST2_scalars}

We consider spinning Calogero-Sutherland Hamiltonians that arise when one of 
the representations $\rho_L,\rho_R$ is trivial. For concreteness, let $\rho_L$ be 
the trivial representation. The potential in the Schr\"odinger operator the simplifies 
to
\begin{equation}
    V = -\frac{D^2_+}{2\sinh^2(t_1+t_2)} - \frac{D^2_-}{2\sinh^2(t_1-t_2)}+ \sum_{a=3}^d 
    \left(\frac{M^2_{1a}}{\sinh^2 t_1} + \frac{M_{2a}^2}{\sinh^2 t_2}\right) - \frac12 M^{ab}M_{ab}\ .
\end{equation}
We assume that $\rho_R$ is a mixed symmetry tensor $(l,\ell)$ of depth two of 
the rotation group. Thus, the generators may be realised as differential 
operators
\begin{align}
    & iM_{12} \to \rho_R^\ast(-iM_{12}) = z^a \partial_a -l, \quad M_{ab} \to z_b \partial_{z^a} - z_a \partial_{z^b} + w_b \partial_{w^a} - w_a \partial_{w^b},\\[2mm]
    & iM_{1a} \to \frac12 \left((1+z^2)\partial_{z^a} - 2z_a (z^b \partial_{z^b} - l) + 2z^b(w_b \partial_{w^a} - w_a \partial_{w^b})\right),\\[2mm] 
    & M_{2a} \to\frac12 \left((1-z^2)\partial_{z^a} + 2z_a (z^b \partial_{z^b} - l) - 2z^b(w_b \partial_{w^a} - w_a \partial_{w^b})\right),
\end{align}
that act on polynomial functions of $z^a$ and $w^a$. In the following, to simplify notation, we will write the differential operators $\rho_R^\ast(-M_{ab})$ simply as $M_{ab}$. The invariance condition
\eqref{m-invariance} reads $M_{ab}f=0$ and is solved by functions of the variables
\begin{equation}
    X = z^a z_a, \quad W = w^a w_a, \quad Y = z^a w_a\ .
\end{equation}
To get to the carrier space of $\rho_R^\ast$, we are further required to impose the 
homogeneity $Y\partial_Y f=\ell f$ and restrict to the 
lightcone $\{W=0\}$. Individual pieces of the Hamiltonian 
\eqref{universal-Hamiltonian} restrict to well-defined operators on such 
functions. Introducing the operator
\begin{align}
    L_{l,\ell}(X) &= -X(1-X)^2 \partial_X^2 
    - \left(\ell(1-X)-2(1-l)X +
    \frac{d-2}{2}(1+X)\right)(1-X)\partial_X\nonumber\\[2mm] 
    & + \left(1-l-\frac{d-2}{2}\right)(\ell(1-X) + lX) 
    - \frac{l(d-2)}{2},
\end{align}
we have
\begin{equation}
    D = 2b,\quad iM_{12} = 2X\partial_X + \ell - l, \quad M_{1a}M_{1a} = L(X), \quad M_{2a}M_{2a} = L(-X)\ .
\end{equation}
Therefore, the Hamiltonian reads
\begin{align}
     H^{(b)}_{l,\ell} &= \partial_{t_1}^2 + \partial_{t_2}^2 +\frac{1 - (2b+\ell-l 
     + 2X\partial_X)^2}{2\sinh^2(t_1+t_2)} + \frac{1 -(2b-\ell+l-2X\partial_X)^2}{2\sinh^2(t_1-t_2)}\nonumber\\[2mm]  
     & + \frac{L_{l,\ell}(X)-\frac14(d-2)(d-4)}{\sinh^2 t_1} +
     \frac{L_{l,\ell}(-X)- \frac14 (d-2)(d-4)}{\sinh^2 t_2} - \frac{d^2-2d+2}{2}\ .
\label{eq:HCSMST3scalars} 
\end{align}
Note that the Hamiltonian acts on three variables, $t_1, t_2$ and $X$. We will compare it 
to the second order differential operator \eqref{eq:D2MSTlimit} we obtained in the previous
section when studying the OPE limit in the central intermediate link of the six-point function. 

\subsubsection{Two STTs and two scalars}
\label{sect:CS_TwoSTT}
Let us now address the second case in which the left and right representations of the 
rotation group are both symmetric traceless tensors. This leads to the Calogero-Sutherland
Hamiltonian as a differential operator in five variables. In the universal Calogero-Sutherland
Hamiltonian, we are required to make substitutions
\begin{align}
    & iM'_{12}\to\rho_L(iM_{12}) = -z'^a \partial'_a +l', \quad M'_{ab} \to z'_a \partial'_b - z'_b \partial'_a,\\[2mm]
    & iM'_{1a} \to -\frac12 \left((1+z'^2)\partial'_a  - 2z'_a (z'^b \partial'_b - l')\right), 
    \quad M'_{2a} \to -\frac12 \left( (1-z'^2)\partial'_a + 2z'_a (z'^b \partial'_b - l')\right),\\[2mm]
    & iM_{12} \to \rho_R^\ast(-iM_{12}) = z^a \partial_a - l, \quad M_{ab} \to z_b \partial_a  - z_a \partial_b,\\[2mm]
    & iM_{1a} \to \frac12 \left((1+z^2)\partial_a - 2z_a (z^b \partial_b - l)\right), 
    \quad M_{2a} \to\frac12 \left( (1-z^2)\partial_a + 2z_a (z^b \partial_b - l)\right)\ .
\end{align}
The invariance condition $(M'_{ab}-M_{ab})f=0$ then means that $f$ 
depends only on the scalar products
\begin{equation}
    X_R = z^a z_a, \quad X_L = z'^a z'_a, \quad Y = z^a z'_a\ .
\end{equation}
Individual pieces of the Hamiltonian commute with the constraints, so they reduce to operators 
in the variables $(X_R,X_L,Y)$. To write them down, we introduce
\begin{align}
    & L_1(X_R,X_L,Y) = \frac14\Big( -4X_R(1-X_R)^2\partial_{X_R}^2 - 4Y(1-X_R)^2\partial_{X_R}\partial_Y + (4Y^2-(1+X_R)^2 X_L)\partial_Y^2\nonumber\\[2mm]
    & + 2(1-X_R) (4(1-l)X_R - (d-2)(1+X_R))\partial_{X_R}\\[2mm]
    & - 2Y ((-2l-d+4)X_R + 2l - d+2) \partial_Y + 2l(2(1-l)X_R - (d-2)(1+X_R))\Big)\nonumber,
\end{align}
and
\begin{align}
    &   L_2(X_R,X_L,Y) = -\frac14\Big(-4Y(1-X_R)(1-X_L) \partial_{X_R}\partial_{X_L} - 2(1-X_R)(X_R(1+X_L) - 2Y^2)\partial_{X_R}\partial_Y\nonumber\\[2mm]
    &  - 2(1-X_L)(X_L(1+X_R)- 2Y^2)\partial_{X_L}\partial_Y + (-4Y^3 + Y(-1+X_R+X_L+3 X_R X_L))\partial_Y^2\\[2mm]
    &  + \left((2-d)(1+X_R)(1+X_L)+2(1-l)X_R(1+X_L)+2(1-l')X_L(1+X_R)-4(1-l-l')Y^2\right)\partial_Y\nonumber \\[2mm]
    &  - 4lY(1-X_L)\partial_{X_L} - 4l'Y(1-X_R)\partial_{X_R} - 4ll' Y\Big)\nonumber\ .
\end{align}
Then we have
\begin{align}
    & M_{1a} M_{1a} =  L_1(X_R,X_L,Y), \quad M_{2a}M_{2a} = L_1(-X_R,-X_L,-Y),\\[2mm]
    & M'_{1a} M_{1a} = L_2(X_R,X_L,Y), \quad M'_{2a} M_{2a} = L_2(-X_R,-X_L,-Y)\ .
\end{align}
Clearly, the operators $M'_{1a}M'_{1a}$ and $M'_{2a}M'_{2a}$ are obtained from 
$M_{1a}M_{1a}$ and $M_{2a}M_{2a}$ by exchanging $X_R$ and $X_L$. 
Together with 
\begin{align}
    & iM_{12} = 2X \partial_{X_R} + Y\partial_Y - l, \quad  iM'_{12} = -2X_L \partial_{X_L} - Y\partial_Y + l',\\[2mm]
    & \frac12 M^{ab} M_{ab} = \left(X_R X_L - Y^2\right)\partial_Y^2 - (d-3)Y\partial_Y\ .
\end{align}
these expressions are substituted in the formula \eqref{universal-Hamiltonian} 
for the universal Hamiltonian to give the appropriate Calogero-Sutherland model 
that characterises four-point blocks 
with two scalars and two STTs. It will be compared with the Hamiltonian we 
obtained in the previous section when we studied the double sided OPE limit 
of the six-point function. 

\subsection{Mapping between OPE-reduced operators and Calogero-Sutherland form}

In this section we will see how, using the leading behaviours of six-point blocks 
spelled out in section~\ref{sect:OPElimits} under the various OPE limits, it is 
possible to recover the spinning four-point Casimir equations from the previous 
subsection. Our strategy is to map the differential equations we obtained when 
discussing OPE limits into a certain ``standard form'' of a quantum mechanical 
Hamiltonian, which can then be identified with one of the spinning Calogero-Sutherland 
Hamiltonians constructed above. As a rule of thumb, we will modify the 
second-order derivatives by performing a change of variables; all first-order 
derivatives can instead be modified without affecting the second-order ones by 
``extracting'' a certain function of the cross ratios from the target function. After this second operation, the new differential operator $\mathcal{D}'$ is 
related to the original one by conjugation by the factor $\omega$
\begin{equation}\label{eq:conjugation_generic}
    \psi =\omega \psi' \qquad \Longrightarrow \qquad \mathcal{D}'=\omega^{-1}
    \mathcal{D}\omega\ .
\end{equation}
Let us now describe these steps in some more detail for the two cases we have analysed in the previous section.

\subsubsection{One \texorpdfstring{MST$_2$}{MST2} and three scalars}
As a first step, it is possible to employ the change of variables used in \cite{Isachenkov:2016gim} in order to map the 
second-order derivatives of $\mathcal{D}_a^2$ in $z_1$ and $\bar{z}_1$ 
to one-dimensional kinetic terms. More precisely, we make the 
following change of variables
\begin{equation}
    z_1, \bar{z}_1 \qquad \longrightarrow \qquad \begin{cases}
        t_1=\mathrm{i} \left[\arcsin\left(\frac{1}{\sqrt{z_1}}\right)
        + \arcsin\left(\frac{1}{\sqrt{\bar{z}_1}}\right)\right],   \\[2mm]
        t_2=\mathrm{i} \left[\arcsin\left(\frac{1}{\sqrt{z_1}}\right)
        - \arcsin\left(\frac{1}{\sqrt{\bar{z}_1}}\right)\right]  ,
    \end{cases}
\end{equation}
which leads to the anticipated transformation in the quadratic Casimir
\begin{equation}
        2z_1^2(1-z_1) \partial_{z_1}^2+2\bar{z}_1^2(1-\bar{z}_1)\partial_{z_1}^2 
        \qquad \longrightarrow \qquad  \partial_{t_1}^2+\partial_{t_2}^2\,.
\end{equation}
Secondly, we wish to eliminate first order derivatives with respect to $t_1$ 
and $t_2$. We can do so through conjugation of the type in eq.\ 
\eqref{eq:conjugation_generic}, with the factor
\begin{multline}
    \omega(t_1,t_2,w_1)=\left(\sinh t_1 \sinh t_2\right){}^{1-\frac{d}{2}} \left(\cosh t_1-\cosh t_2\right){}^{-a-b-\frac{1}{2}} \left(\cosh t_1+\cosh t_2\right){}^{a+b-\frac{1}{2}}\\[2mm]
    \left(\frac{\sinh t_1+\sinh t_2}{\sinh t_1-\sinh t_2}\left(1-w_1\right){}^2\right)^{\frac{l_b-\ell _b}{2}}\,.
\end{multline}
We now wish to eliminate the mixed derivatives  $\partial_{t_i}\partial_{w_1}$, 
providing this way a partial decoupling of the internal leg and vertex degrees 
of freedom. This can be achieved by the change of variables
\begin{equation}\label{w-in-terms-of-X}
  w_1 \qquad \longrightarrow \qquad   X=\frac{\sinh t_1-\sinh t_2}
  {\sinh t_1+\sinh t_2}\frac{w_1}{1-w_1}\,.
\end{equation}
The operator produced at the end of this procedure is of the spinning Calogero-Sutherland 
form, and corresponds precisely to the operator we spelled out in the previous 
subsection~\ref{sect:CS_MST2_scalars}, see eq.\ \eqref{eq:HCSMST3scalars}.

\subsubsection{Two STTs and two scalars}
As we did in the previous subsection, the first step to map this operator to 
a quantum mechanical Hamiltonian is to make two derivatives become one-dimensional 
kinetic terms, which is done by transforming
\begin{equation}
    z_2, \bar{z}_2 \qquad \longrightarrow \qquad \begin{cases}
        t_1=\mathrm{i} \left[\arcsin\left(\frac{1}{\sqrt{z_2}}\right)
        + \arcsin\left(\frac{1}{\sqrt{\bar{z}_2}}\right)\right],   \\[2mm] 
        t_2=\mathrm{i} \left[\arcsin\left(\frac{1}{\sqrt{z_2}}\right)-
        \arcsin\left(\frac{1}{\sqrt{\bar{z}_2}}\right)\right]  ,
    \end{cases}
\end{equation}
which leads to 
\begin{equation}
        2z_2^2(1-z_2) \partial_{z_2}^2+2\bar{z}_2^2(1-\bar{z}_2)\partial_{z_2}^2 
        \qquad \longrightarrow \qquad  \partial_{t_1}^2+\partial_{t_2}^2\,.
\end{equation}
As a second step, it is possible to remove the first order derivatives in $t_1$ and $t_2$ by 
conjugating with
\begin{multline}
    \omega(t_1,t_2,w_1,w_2,\Upsilon)=
   \left(\sinh t_1\sinh t_2\right)^{1-\frac{d}{2}} \left(\cosh t_1-\cosh t_2\right)^{- a- b-\frac{1}{2}} \left(\cosh t_1+\cosh t_2\right){}^{a+b-\frac{1}{2}}\\[2mm]
    \left(\frac{\sinh t_1-\sinh t_2}{\sinh t_1+\sinh t_2}w_1^2\right)^{\frac{l_a}{2}}\left(\frac{\sinh t_1+\sinh t_2}{\sinh t_1-\sinh t_2}\left(1-w_2\right)^2\right)^{\frac{l_b}{2}}\,.
\end{multline}
Finally, we can eliminate all mixed derivatives that involve the $t_i$ 
coordinates by taking the change of variables
\begin{equation}\label{w1-to-XL}
\begin{gathered}
    X_L=\frac{\sinh t_1+\sinh t_2}{\sinh t_1-\sinh t_2}\frac{1-w_1}{w_1}\,, 
    \qquad X_R=\frac{\sinh t_1-\sinh t_2}{\sinh t_1+\sinh t_2}\frac{w_2}{1-w_2}\,,\\[2mm]
    Y=-\sqrt{\frac{(1-w_1)w_2}{(1-w_2)w_1}}\frac{\sinh t_1 \sinh t_2}{\cosh t_1 - 
    \cosh t_2}\Upsilon\,.
    \end{gathered}
\end{equation}
The operator obtained at the end of this procedure corresponds precisely to the one 
we constructed in subsection~\ref{sect:CS_TwoSTT}.

\section{Conclusions and outlook}

The central goal of this work was to introduce a new set of multipoint cross ratios 
for correlation functions in higher dimensional CFT that are well adapted to taking 
OPE limits in the intermediate channels. We succeeded to do so for comb channel 
diagrams of scalar $N$-point functions in $d=3,4$. The strategy we have used here 
lends itself to several fairly obvious extensions. In particular, given the nice 
geometrical interpretation of the new cross ratios we described in section~\ref{sect:OPElimits}, it 
seems obvious how to extend the construction of cross ratios to higher dimension
and a higher number $N$ of insertion points. Recall that every time we increase the 
dimensions $d$, we obtain one new type of cross ratio, either for vertices or for 
links. The most complicated such variable we constructed here was the six-point 
invariant $\Upsilon$ which initially appears for six-point functions in $d=4$. 
When be pass to $d=5$, one can have vertices that carry two variables. This 
appears starting from $N=7$. We expect that this new variable can be related 
to the angle between two 4-dimensional subspaces in $d=5$ dimensions in a way 
that preserves all of the nice properties we described above. 

We have verified the factorisation of multipoint blocks in the OPE limit only 
for $N=6$. Given that in $d=4$ there are no new types of cross ratios as we 
increase the number of cross ratios we conjecture that our factorisation 
statement does extend to any number of insertion points for comb channel 
blocks in $d=4$. Of course, it would be reassuring to prove this statement. 
Further in this direction, it would be interesting to study OPE limits and 
the associated factorisation for higher $N = d+2 > 6$. Recall from our 
discussion in section~\ref{sect:OPE_embedding_space} that each time we increase $d/N$, one new 
polynomial cross ratio appears which involves all $d+2$ insertion points. 
Given the nice geometric interpretation of our OPE variables for $d \leq 4$ 
as angles between hyperplanes of increasing dimension, it is tempting 
generalise the very same construction to higher $d$. In this way it is 
rather easy to construct a complete set of variables that has a very 
good chance to be well adapted to taking OPE limits on intermediate 
exchange channels with more than three cross ratios. Still this should 
be verified at least in a few more examples. Finally, it would certainly 
be very rewarding go beyond the comb channel topology and to construct 
OPE cross ratios for other OPE channels, such as e.g. the snowflake 
channel. 

In this work we have seen how one can recover lower point blocks of 
spinning fields from $N$-point blocks of scalars. In particular, by 
taking the appropriate OPE limits it is possible to recover spinning 
three-point and four-point blocks. Ultimately, it would be important 
to go into the opposite direction, i.e.\ to learn how to build 
multipoint blocks of scalar fields from their three and four-point 
constituents. Note that the latter are reasonably well understood, 
while multipoint blocks in dimensions $d > 2$ remain very 
difficult to evaluate. We have made some partial steps in this 
direction recently in the case of a two variable system that is 
obtained from the 5-point system by taking the full OPE limit on 
one side and a lightcone limit on the other. The resulting system  
of differential operators, which contains a vertex and the lightcone 
limit of a Casimir operator, can indeed be solved in terms of the 
solution of the two subsystems. We plan to return to this and 
other examples in the near future. 

In section~\ref{sect:CSmodels} of this work we have used some recent progress 
concerning a universal construction of Casimir operators for spinning four-point 
blocks. The full derivation of our formulas along with some other applications of 
these developments, e.g.\ to boundary and defect CFT, will be discussed in much 
more detail in the upcoming paper \cite{Buricnew}. The universal construction of 
the Casimir operators also hints at the existence of a universal power series 
solutions based on so-called Harish-Chandra coefficients. Upon replacing these 
coefficients by concrete representation operators, one should be able to obtain 
a series expansion for spinning blocks, at least in principle. 

Recently, the lightcone bootstrap for multipoint blocks was initiated in 
\cite{Vieira:2020xfx}. In the context of the six-point snowflake channel
this program has began to produce interesting new insights into the 
large spin behaviour of three-point couplings involving two and three 
double twist operators \cite{Antunes:2021kmm}. Similarly, adapting the 
lightcone bootstrap to six-point functions in the comb channel is expected 
to provide access to triple trace operators. But for the moment, the 
relevant comb channel blocks are not known, even in the lightcone 
limit. It seems likely that an analysis similar to the one we have 
presented above can overcome this difficulty and furnish sufficient 
control over such comb-channel lightcone blocks.  
\medskip 

\noindent 
\textbf{Acknowledgements:} We are grateful to Gleb Arutyunov, Luke Corcoran, 
Pavel Etingof, Aleix Gimenez-Grau, Mikhail Isachenkov, Apratim Kaviraj, 
Madalena Lemos, Pedro Liendo, Nikolai Reshetikhin, Junchen Rong, Evgeny Sobko, Jasper 
Stokman, Joerg Teschner and Benoît 
Vicedo for useful discussions. This project received funding from the 
German Research Foundation DFG under Germany’s Excellence Strategy -- EXC 
2121 Quantum Universe -- 390833306 and from the European Union’s Horizon 
2020 research and innovation programme under the MSC grant agreement 
No.764850 “SAGEX”. S.L. is supported by Dr. Max R\"ossler, the Walter Haefner Foundation and the ETH Z\"urich Foundation. I.B. is funded by a research grant under the project H2020 ERC STG 2017 G.A. 758903 "CFT-MAP".

\appendix

\section{Euclidean conformal group}
\label{app:Eucl_conf_group}
Here we spell out our conventions for the conformal group $SO(d+1,1)$ and its 
Lie algebra and state some identities valid in the vector representation. 
The non-vanishing Lie brackets in $\mathfrak{so}(d+1,1)$ read
\begin{align}
    & [M_{\mu\nu},P_\rho] = \delta_{\nu\rho} P_\mu 
    - \delta_{\mu\rho} P_\nu,\quad [M_{\mu\nu},K_\rho] = 
    \delta_{\nu\rho} K_\mu - \delta_{\mu\rho} K_\nu,\\[2mm]
    & [M_{\mu\nu},M_{\rho\sigma}] = \delta_{\nu\rho} M_{\mu\sigma} 
    - \delta_{\mu\rho} M_{\nu\sigma} + \delta_{\nu\sigma} M_{\rho\mu} 
    - \delta_{\mu\sigma} M_{\rho\nu},\\[2mm]
    & [D,P_\mu] = P_\mu,\quad [D,K_\mu]=-K_\mu,\quad [K_\mu,P_\nu] 
    = 2(M_{\mu\nu} - \delta_{\mu\nu}D)\ .
\end{align}
Indices $\mu,\nu$ run from $1$ to $d$. In the Lorentz-like notation, we write the generators as 
$\{L_{AB}\}$, $A,B = -1,0,...,d$. These obey the relations
\begin{equation}
    [L_{AB},L_{CD}] = \eta_{BC} L_{AD} - \eta_{AC} L_{BD} + \eta_{BD} L_{CA} - \eta_{AD} L_{CB},
\end{equation}
where $\eta=\text{diag}(-1,1,...,1)$. The relation between conformal and Lorentz generators reads
\begin{equation}
    L_{-1,0} = D, \quad L_{-1,\mu} = \frac12(P_\mu + K_\mu), \quad L_{0\mu} = \frac12(P_\mu - K_\mu), \quad L_{\mu\nu} = M_{\mu\nu}\ .
\end{equation}
In the $(d+2)$-dimensional vector representation, the Lorentz generators are
\begin{equation}
    L_{AB} = \eta_{AC} E_{CB} - \eta_{BC} E_{CA},
\end{equation}
where $ (E_{AB})_{ij} = \delta_{A i} \delta_{B j}$. 

\subsection{Radial decomposition of the Casimir}

The quadratic Casimir $C_2 = -\frac12 L^{AB} L_{AB}$ can 
be written as
\begin{equation}
    C_2  = L_{-1,0}^2 + L_{-1,1}^2 + L_{-1,2}^2 - L_{01}^2 - L_{02}^2 - 
    L_{12}^2 - L^{-1,a}L_{-1,a} - L^{0a}L_{0a} - L^{1a}L_{1a} - L^{2a}L_{2a} - \frac12 L^{ab}L_{ab}\ .
\end{equation}
Here, indices $a,b$ take values $3,4,...,d$. Using relations
\begin{align}
    & L_{-1,a} = \coth t_1 M_{1a} - \frac{1}{\sinh t_1} M'_{1a}
    \quad L_{0a} = i \left(\coth t_2 M_{2a} - \frac{1}{\sinh t_2} M'_{2a}\right),\\[2mm]
    & L_{01}\mp iL_{-1,2} = -\coth(t_1\pm t_2)D_\pm + \frac{D'_\pm}{\sinh(t_1\pm t_2)},\\[2mm]
    & [M_{1a},M'_{1a}] = -\sinh t_1 H_1, \quad [M_{2a},M'_{2a}] = 
    -\sinh t_2 H_2,\quad [D_\pm,D'_\pm] = 2 \sinh(t_1\pm t_2) (H_1 \pm H_2),
\end{align}
we obtain
\begin{align*}
    & - L^{-1,a}L_{-1,a} - L^{1a}L_{1a} = \frac{M'_{1a}M'_{1a}-2\cosh t_1 
    M'_{1a}M_{1a}+M_{1a}M_{1a}}{\sinh^2 t_1} + (d-2)\coth t_1 H_1,\\[2mm]
    & - L^{0a}L_{0a} - L^{2a}L_{2a} = \frac{M'_{2a}M'_{2a}-2\cosh t_2 
    M'_{2a}M_{2a}+M_{2a}M_{2a}}{\sinh^2 t_2} + (d-2)\coth t_2 H_2,
\end{align*}
and 
\begin{align*}
    L_{-1,0}^2 - L_{12}^2 + L_{-1,2}^2 - L_{01}^2 & = - \frac{D'^2_+ 
    - 2\cosh(t_1+t_2) D'_+ D_+ + D^2_+}{2\sinh^2(t_1 + t_2)} 
    - \frac{D'^2_- - 2\cosh(t_1-t_2) D'_- D_- + D^2_-}{2\sinh^2(t_1 - t_2)}\\[2mm]
    & + \coth(t_1+t_2) (H_1+H_2) + \coth(t_1-t_2)(H_1-H_2)\ .
\end{align*}
Adding these terms gives the radial decomposition used in the main text.

\section{Construction of a six-point conformal frame}
\label{app:conformal_frames}
In this section, we construct a conformal frame for the six-point function by 
appending $x_6$ to the conformal frame of the $(12345)$ five-point function, 
namely:
 \begin{align}
 x_1 = \varrho_1 \vec{n}(\theta_1,0,0),\quad 
 x_2 = 0, \quad
 x_3 = \infty, \quad 
 x_4 = \vec{e}_1, \quad
 x_5 = \vec{e}_1 - \varrho_2 \vec{n}(\theta_2,\phi_1,0),
\label{constraints_frame}
\end{align}
where we parameterise unit vectors in $S^4$ as
\begin{equation}
\vec{n}(\theta,\phi,\varphi) := \cos\theta\, \vec{e}_1 + \sin\theta\big\{\cos\phi\,\vec{e}_2+ \sin\phi \left( \cos\varphi\,\vec{e}_3 + \sin\varphi\,\vec{e}_4\right) \big\} = e^{\varphi M_{34}} e^{\phi M_{23}} e^{\theta M_{12}} \vec{e}_1.
\end{equation}
It will also be useful to define the rotation matrices
\begin{equation}
R(\theta,\phi,\varphi):= e^{-\theta M_{12}}e^{-\phi M_{23}}e^{-\varphi M_{31}} \Longrightarrow \vec{n}(\theta,\phi,\varphi)=R(\theta,\phi,\varphi)^{-1} \vec{e}_1.
\end{equation}
Finally, we parameterise cross ratios as in \eqref{eq:6ptcrgeom},
\begin{equation}
z_r := \varrho_r e^{\mathrm{i} \theta_r}, \quad \bar{z}_r :=\varrho_r e^{-\mathrm{i} \theta_r}, \quad w_s := \sin^2 \frac{\phi_s}{2}, \quad \Upsilon := \pm \mathrm{i} \frac{\cos \varphi}{\sin \theta_2}.
\end{equation}
To understand how $x_6$ depends on the cross ratios, we compute a distinguished vector in this frame:
\begin{equation}
\psi_{56} := ( x_{45}^{-1}-x_{46}^{-1})^{-1} \in\mathbb{R}^4_{1234}, 
\end{equation}
where $x^{-1}:=x/x^2$ denotes the image of the vector $x$ under conformal inversion. Note that we implicitly used the residual $\mathrm{SO}(d-4)$ symmetry preserving \eqref{constraints_frame} to fix a gauge where $x_6\in\mathrm{Span}(\vec{e}_1,\vec{e}_2,\vec{e}_3,\vec{e}_4)$. In Euclidean signature, we can parameterise the latter by its norm and its unit vector on the $S^4$, which we write as
\begin{equation}
\psi_{56} = \abs{\psi_{56}}\,  \hat{\psi}_{56}, \quad \abs{\psi_{56}} = \varrho_2 \varrho_3^{-1}, 
\end{equation}
Then the unit vector $\hat{\psi}_{56}$ is determined by three equations:
\begin{equation}
 \hat{\psi}_{56} \cdot \hat{x}_{45}=\frac{1+u_2-v_2}{2 \varrho_2},\quad \hat{\psi}_{56} \cdot x_4 = \frac{\mathcal{U}_2^{(5)}}{2 \varrho_2\varrho_3}, \quad \hat{\psi}_{56} \cdot \hat{x}_1 = \frac{\mathcal{U}_1^{(6)}}{2 \varrho_1 \varrho_2 \varrho_3},
 \label{psi56angles}
\end{equation}
where the $\mathcal{U}_r^{(m)}$ are polynomials in the polynomial cross ratios:
\begin{align}
&\mathcal{U}_r^{(5)} := 1-v_r-v_{r+1}+U_r^{(5)}, \quad \mathcal{U}_1^{(6)} := (1-v_1)(1-v_3)-v_2 + U_1^{(5)}+U_2^{(5)}-U_1^{(6)}. \label{curlyUs} 
\end{align}
Using the change of variables \eqref{w2def} and \eqref{eq:Upsilondef}, we can express the scalar products of \eqref{psi56angles} in terms of the angle cross ratios $(\theta_s,\phi_r,\varphi)$:
\begin{align}
\frac{\mathcal{U}_1^{(5)}}{2\varrho_2\varrho_3} =& \cos\theta_2 \cos\theta_3 + \sin\theta_2\sin\theta_3\cos\phi_2 \\
\frac{\mathcal{U}_1^{(6)}}{2 \varrho_1 \varrho_2 \varrho_3} 
=&\cos\theta_1 \cos\theta_2\cos\theta_3  \nonumber + \cos\theta_1 \sin\theta_2\cos\phi_2\sin\theta_3 \nonumber+\sin\theta_1 \sin\theta_2\cos\phi_1\cos\theta_3  \nonumber\\[2mm]
	     &-\sin\theta_1\left(\cos\theta_2\cos\phi_1\cos\phi_2 + \sin\phi_1\sin\phi_2\cos\varphi\right)\sin\theta_3.
\end{align}
Given that $x_4 = \vec{e}_1$, $\hat{x}_1\in \mathrm{Span}(\vec{e}_1,\vec{e}_2)$, $\hat{x}_{45} \in\mathrm{Span}(\vec{e}_1,\vec{e}_2,\vec{e}_3)$, we can recursively compute the components of $\hat{\psi}_{56}$ as
\begin{align*}
&\hat{\psi}_{56} \cdot x_4 = \hat{\psi}_{56} \cdot \vec{e}_1 \Longrightarrow \hat{\psi}_{56}^{1}, \\[2mm]
& \hat{\psi}_{56} \cdot \hat{x}_1 = \hat{\psi}_{56} \cdot \vec{n}(\theta_1,0,0) \Longrightarrow \hat{\psi}_{56}^{2}, \\[2mm]
&\hat{\psi}_{56} \cdot \hat{x}_{45} = \hat{\psi}_{56} \cdot \vec{n}(\theta_2,\phi_1,0) \Longrightarrow \pm \hat{\psi}_{56}^{3}, \\[2mm]
&(\hat{\psi}_{56}^{1})^2 +(\hat{\psi}_{56}^{2})^2+(\hat{\psi}_{56}^{3})^2+(\hat{\psi}_{56}^{4})^2=1 \Longrightarrow \pm \hat{\psi}_{56}^{4}.
\end{align*}
There is a sign indeterminacy in the third step coming from the convention for $\Upsilon \propto \pm \cos\varphi$, and there is  also a sign indeterminacy in the last step coming from the two solutions to the quadratic equation. We compute the solution to each of these equations and find
\begin{equation}
\hat{\psi}_{56} = R(\theta_2,\phi_1,0)^{-1} \vec{n}(-\theta_3,\pm\phi_2,\pm\varphi) \stackrel{\mathrm{choice}}{=} R(\theta_2,\phi_1,0)^{-1}\vec{n}(-\theta_3,\phi_2,\varphi).
\end{equation}

Now, we can obtain $x_6$ in the conformal frame by a conformal transformation of $\psi_{56}$,
\begin{equation}
\psi_{56} = (x_{64}^{-1}-x_{54}^{-1})^{-1} = e^{-x_{45}^{-1}\cdot K} e^{x_4\cdot P} \cdot x_6 = e^{-\varrho_2^{-1} \vec{n}(\theta_2,\phi_1,0)\cdot K} e^{P_1} x_6.
\label{psi56_to_x6}
\end{equation}
After simplifying and inverting the conformal transformation in \eqref{psi56_to_x6}, we then obtain
\begin{equation}
x_6 = e^{-P_1} \varrho_2^{-D} R(\theta_2,\phi_1,0)^{-1} \mathcal{I} \, e^{P_1} \cdot  \varrho_3 n(-\theta_3,\phi_2,\varphi),
\end{equation}
where $\mathcal{I}: x\mapsto x^{-1}$ is conformal inversion. To better understand the meaning of these conformal transformations, let's take a closer look at the conformal group element
\begin{equation}
g \equiv g(\varrho_2,\theta_2,\phi_1) = e^{-P_1} \varrho_2^{-D} R(\theta_2,\phi_1,0)^{-1}\mathcal{I} \, e^{P_1}.
\end{equation} 
Its inverse acts as
\begin{equation}
g^{-1}: x \mapsto \varrho_2^{-1} R(\theta_2,\phi_1) (x-\vec{e}_1)^{-1}+\vec{e}_1, 
\end{equation}
such that when $g^{-1}$ acts on the points of the original conformal frame, the images are given by
\begin{equation}
g^{-1}(x_6) = \varrho_3 \vec{n}(-\theta_3,\phi_2,\varphi), \quad g^{-1}(x_5) = 0, \quad g^{-1}(x_4) = \infty,  \quad g^{-1}(x_3)= \vec{e}_1.
\end{equation}

This suggests a general method to characterise the comb channel conformal frame of $N>6$ points in $d=4$, which depends on the cross ratios $(\varrho_r,\theta_r)_{r=1}^{N-3}$, $(\phi_s)_{s=1}^{N-4}$ and $(\varphi_r)_{r=2}^{N-4}$ defined in \eqref{eq:OPEcrgeom}. First, define the conformal transformation
\begin{equation}
h^{-1}(\varrho_2,\theta_2,\phi_1,0,\varphi_2) := e^{- \varphi_2 M_{34}}\sigma_1 \, e^{P_1} g^{-1}(\varrho_2,\theta_2,\phi_1) =\varrho_2^{-D}\mathcal{I} \sigma_1  e^{- \varphi_2 M_{zt}} R(\theta_2,\phi_1)  e^{P_1},
\end{equation}
where $\sigma_1: (x^1,x^2,x^3,x^4) \mapsto (-x^1,x^2,x^3,x^4)$ is a reflection along the hyperplane orthogonal to $\vec{e}_1$. Its action on a point is given by
\begin{equation}
    h^{-1}(x) = \varrho_2 \sigma_1 e^{- \varphi_2 M_{34}} R(\theta_2,\phi_1) (x-\vec{e}_1)^{-1}.
\end{equation}
From the previous discussion, we determined that this conformal transformation acts on the six-point conformal frame as follows:
\begin{equation}
h^{-1}(x_2) = \varrho_2 \vec{n}(\theta_2,0,0), \quad h^{-1}(x_3) = 0, \quad h^{-1}(x_4) = \infty, \quad h^{-1}(x_5) = \vec{e}_1, \quad h^{-1}(x_6) = \vec{e}_1 - \varrho_3 \vec{n}(\theta_3,\phi_2,0).
\end{equation}
Thus, $h^{-1}$ shifts the framing from the constraints \eqref{constraints_frame} on $x_i$, to the same constraints on $x_{i+1}$, $i=1,\dots,5$. We can similarly express the seventh point as
\begin{equation}
h^{-1}(x_7) = h' \left( \vec{e}_1 - \varrho_4 \vec{n}(\theta_4,\phi_3,0)\right),  
\end{equation}
where $h'\equiv h(\varrho_3,\theta_3,\phi_2,\varphi_2,\varphi_3)$ is now uniquely defined by
\begin{align*}
& h'^{-1}(0) = \varrho_3 \vec{n}(\theta_3,0,0), \quad h'^{-1}(\infty) = 0, 
\quad h'^{-1}(e_x) = \infty,\\[2mm]
& h'^{-1}(h^{-1}(x_6)) = \vec{e}_1, \quad h'^{-1}(h^{-1}(x_7)) = 
\vec{e}_1 - \varrho_4 \vec{n}(\theta_4,\phi_3,0). 
\end{align*}
A quick comparison with the action of $h^{-1}$ on $x_2,\dots,x_6$ implies that 
\begin{equation}
h'^{-1} := \varrho_3^{-D} \mathcal{I} \sigma_1 e^{- \varphi_3 M_{34}} R(\theta_3,\phi_2,\varphi_2)^{-1}  e^{P_1}.
\end{equation}
We can then iterate this procedure until reaching $x_N$. More specifically, the frame will be given by
 \begin{align*}
& x_1 = \varrho_1 \vec{n}(\theta_1,0,0), \\[2mm]
& x_2 = 0, \\[2mm]
& x_3 = \infty, \\[2mm]
& x_4 = \vec{e}_1, \\[2mm]
& x_5 = \vec{e}_1 - \varrho_2 \vec{n}(\theta_2,\phi_1,0), \\[2mm]
& x_6 = h(\varrho_2,\theta_2,\phi_1,\varphi_1,0)(\vec{e}_1 - \varrho_3 \vec{n}(\theta_3,\phi_2,0)) \\[2mm]
& x_7 = h(\varrho_2,\theta_2,\phi_1,\varphi_1,0) \circ h(\varrho_3,\theta_3,\phi_2,\varphi_1,\varphi_2)(\vec{e}_1 - \varrho_4 \vec{n}(\theta_4,\phi_3,0)),\\[2mm]
& \dots \dots \dots \\[2mm]
& x_N = h(\varrho_2,\theta_2,\phi_1,\varphi_1,0) \circ  \prod_{r=3}^{N-4} h(\varrho_{r},\theta_{r},\phi_{r-1},\varphi_{r-1},\varphi_r)  \left(\vec{e}_1 -\varrho_N \vec{n}(\theta_N,\phi_{N-1},\varphi_{N-2},0)\right),
\end{align*}
where
\begin{equation}
h^{-1}(\varrho_{r},\theta_{r},\phi_{r-1},\varphi_{r-1},\varphi_r):= \varrho_r^{-D} \mathcal{I} \sigma_1 e^{- \varphi_r M_{34}} R(\theta_r,\phi_{r-1},\varphi_{r-1})^{-1}  e^{P_1}.
\end{equation}
 The action of this conformal group element on points is then given by 
\begin{equation}
h^{-1}(\varrho_{r},\theta_{r},\phi_{r-1},\varphi_{r-1},\varphi_r): x \mapsto  \varrho_{r} \sigma_1 e^{-\varphi_{r} M_{34}}R(\theta_{r},\phi_{r-1},\varphi_{r-1})^{-1} (x-\vec{e}_1)^{-1}.
\end{equation}

\section{Middle leg OPE limit in embedding space}
\label{app:bOPE_embedding_space}
In the six-point function, the limit $\bar{z}_2 \rightarrow 0$ at the middle leg $b$ can be lifted to embedding space as
\begin{equation}
X_{45},X_{46},X_{56} \rightarrow 0, \quad \frac{X_{45}}{X_{46}}, \frac{X_{56}}{X_{46}} = \mathrm{finite}. \label{zb2OPElift}
\end{equation}

In other words, all distances between the three points $x_4,x_5$ and $x_6$ vanish at the same rate in spacetime.  By first quantising around $x_6$ and then mapping to the cylinder, a triplet satisfying \eqref{zb2OPElift} is mapped to past timelike infinity. The infinite distance between $(x_4,x_5,x_6)$ and $(x_1,x_2,x_3)$ in this limit factorises the six-point function into a product of two four-point functions in a manner reminiscent of the cluster decomposition principle. 

To compute this limit in embedding space, it will be useful to define the vector
\begin{align}
Y_5:= (X_4-X_5) - \frac{X_{45}}{X_{46}}(X_4-X_6).
\label{Y2Y5}
\end{align}
In particular, $X_4\wedge Y_5$ is a homogeneous tensor in both $X_4$ and $X_5$. For a spacetime interpretation of this vectors, recall the reduction $X,Z \mapsto x,z$ of a STT to the Poincare patch:
\begin{equation}
X_x = \left(\frac{1+x^2}{2},\frac{1-x^2}{2},x\right)\,, \quad Z_{x,z}=(x \cdot z, - x\cdot z,z).
\end{equation}
If we put all embedding space vectors in this patch, $X_i :=X_{x_i}$, we obtain
\begin{equation}
Y_5 = \frac{x_{56}^2}{x_{46}^2} Z_{x_4,\psi_{56}}, \quad \psi_{56}:=(x_{45}^{-1}-x_{46}^{-1})^{-1}. 
\end{equation}
Note also that $Y_5^2 = -2 \frac{X_{45}X_{56}}{X_{46}} \rightarrow 0$ in the limit \eqref{zb2OPElift}. We now define the full $b\,$OPE limit $z_2,\Upsilon\rightarrow 0$ in embedding space as
\begin{equation}
X_5 = X_4 + \epsilon Z_4, \quad Y_5 = \epsilon \epsilon' W_4, \quad \epsilon,\epsilon' \rightarrow 0. \label{MST_OPE_lim}
\end{equation}
where $Z_4$ and $W_4$ are MST$_2$ polarisation vectors for $X_4$. To make the connection with the prescription of section \ref{sect:OPE_embedding_space} more explicit, we can rewrite the second equation in \eqref{MST_OPE_lim} as
\begin{equation}
\epsilon' W_4 = Z_4' - Z_4, \quad  \epsilon Z_4' := \frac{X_{45}}{X_{46}  }(X_6-X_4). 
\end{equation}
We thereby obtain the same prescription as eq. \eqref{Wmst2} up to projective equivalence, with a rescaling of $Z_4'$ outside of the conventional Poincare patch to simplify computations. Note that the permutation $(4,5,6) \leftrightarrow (3,2,1)$ leads to an identical parameterisation of the $b\,$OPE limit. Thus, to make expressions more symmetric, we also define 
\begin{equation}
    Y_2 := Y_5 \vert_{(4,5,6) \leftrightarrow (3,2,1)}.
\end{equation}
Now, to understand how $\Upsilon$ encodes MST$_2$ transfer along the internal leg $b$, we would like to compute the $b\,$OPE limit of 
\begin{equation}
 \frac{X_{34} (X_3\wedge X_2\wedge Y_2)\cdot (X_4\wedge X_5\wedge Y_5)}{X_{24}^2X_{35}^2} = \mathcal{U}_1^{(5)}\mathcal{U}_2^{(5)}-(1-v_2)\mathcal{U}_1^{(6)}, 
 \label{MST_CR}
\end{equation}
where $\mathcal{U}_r^{(m)}$ are the same functions of the polynomial cross-ratios defined in \eqref{curlyUs}. To relate them to the left hand side of \eqref{MST_CR}, we expressed them in embedding space as follows:
\begin{align}
&\mathcal{U}_1^{(5)} =  \frac{(X_3 \wedge Y_2) \cdot (X_4 \wedge X_5)}{X_{24}X_{35}}, \quad 
\mathcal{U}_2^{(5)} = \frac{(X_3 \wedge X_2) \cdot (X_4 \wedge Y_5)}{X_{24}X_{35}}, \quad \mathcal{U}_1^{(6)} = \frac{(X_3 \wedge Y_2)\cdot(X_4\wedge Y_5)}{X_{24}X_{35}}.
\end{align} 
On the left hand side of \eqref{MST_CR}, the OPE limit \eqref{MST_OPE_lim} is simple to compute: 
\begin{align*}
\mathrm{LHS} =\epsilon^2 \epsilon' \frac{X_{23} U_{4,123} }{X_{13}X_{24}^2X_{34}}+\mathcal{O}(\epsilon^3 \epsilon'), \quad U_{4,123} := (X_4\wedge Z_4\wedge W_4)_{ABC} X_1^AX_2^BX_3^C. 
\end{align*}
Note that $U_{4,123}$ is the unique independent MST$_2$ tensor structure of the four-point function of three scalars and one MST$_2$ field. On the other hand, the right hand side of \eqref{MST_CR} can be written in cross ratio space using
\begin{align}
\mathcal{U}_1^{(5)} &= z_1' z_2+\bar{z}_1' \bar{z}_2, \\[2mm]
\mathcal{U}_2^{(5)} &= z_2z_3'+\bar{z}_2\bar{z}_3'  ,\\[2mm]
\mathcal{U}_1^{(6)} &= z_1' z_2 z_3' + \bar{z}_1' \bar{z}_2 \bar{z}_3' - \left[(z_1-\bar{z}_1) \sqrt{w_1(1-w_1)}\right] (z_2-\bar{z}_2) \Upsilon \left[(z_3-\bar{z}_3) \sqrt{w_2(1-w_2)}\right],
\end{align}
where we defined
\begin{align}
z_1' := w_1 z_1 + (1-w_1) \bar{z}_1, \quad  \bar{z}_1' := w_1 \bar{z}_1 + (1-w_1) z_1,  \\[2mm]
z_3' := w_2 z_3 + (1-w_2) \bar{z}_3, \quad  \bar{z}_3' := w_2 \bar{z}_3 + (1-w_2) z_3.
\end{align}
Taking $\bar{z}_2=0$, we then find
\begin{equation}
 \mathcal{U}_1^{(5)}\mathcal{U}_2^{(5)}-(1-v_2)\mathcal{U}_1^{(6)} =\frac14 \left[(z_1-\bar{z}_1) \sqrt{w_1(1-w_1)}\right] z_2^2\Upsilon  \left[(z_3-\bar{z}_3) \sqrt{w_2(1-w_2)}\right].
\end{equation}
At the same time, using
\begin{equation}
1-v_2 = z_2= \epsilon \frac{J_{4,23}}{X_{24} X_{34}} + \mathcal{O}(\epsilon^2),  \quad
 J_{4,23} := (X_4\wedge Z_4)_{BC} X_2^BX_3^C,
 \label{z2_embedding_space}
\end{equation}
we find that $\Upsilon=\mathcal{O}(\epsilon')$ in the $b\,$OPE limit \eqref{MST_OPE_lim} with leading behaviour
\begin{equation}
\Upsilon = \epsilon' \frac{ U_{4,123} X_{23}X_{34}}{X_{13}J_{4,23}^2} \left[\frac{z_1-\bar{z}_1}{2} \frac{z_3-\bar{z}_3}{2} \sqrt{w_1(1-w_1)} \sqrt{w_2(1-w_2)}\right]^{-1} + \mathcal{O}(\epsilon\epsilon').
\label{Upsilon_embedding_space}
\end{equation}
In particular, if we define 
\begin{equation}
\mathrm{deg} \, X_4^{-\Delta}Z_4^lW_4^{\ell} := [\Delta,l,\ell] \Longrightarrow \mathrm{deg} \, J_{4,23}=[-1,1,0], \quad \mathrm{deg} \,U_{4,123}=[-1,1,1],
\end{equation}
then we find from \eqref{z2_embedding_space} and \eqref{Upsilon_embedding_space} that
\begin{equation}
\mathrm{deg}\,z_2 = [1,1,0], \quad \mathrm{deg}\,\bar{z}_2 = [1,-1,0], , \quad \mathrm{deg}\,\Upsilon = [0,-1,1].
\end{equation}
This fits directly with the asymptotic behaviour \eqref{eq:bOPE_to_4pt} of conformal blocks in the $b\,$OPE limit.

\section{OPE limit factorisation of six-point blocks in one-dimensional CFT}
\label{app:rosenhaus}
Let us consider the case of six-point conformal blocks in the comb channel for 
the $d=1$ case. The conformal blocks for this case have been computed 
in~\cite{Rosenhaus:2018zqn}. To match the convention of that paper, we will 
introduce the three cross ratios
\begin{equation}
    \chi_1=\frac{x_{12}x_{34}}{x_{13}x_{24}}\,, \qquad \chi_2=\frac{x_{23}x_{45}}{x_{24}x_{35}}\,, \qquad \chi_3=\frac{x_{34}x_{56}}{x_{35}x_{46}}\,,
\end{equation}
which make a complete set of independent cross ratios in $d=1$, and we 
rename conformal dimensions as
\begin{equation}
    \Delta_i=h_i\,, \qquad \Delta_a=\mathbbm{h}_1\,, \qquad \Delta_b=\mathbbm{h}_2\,, \qquad \Delta_c=\mathbbm{h}_3\,.
\end{equation}
Note that when reducing to $d=1$, the Gram determinant relations impose 
$z_i=\bar{z}_i$, and that
\begin{equation}
    \chi_i=z_i=\bar{z}_i\,.
\end{equation}
The one-dimensional six-point conformal blocks can then be written as in 
\cite[equation 2.11]{Rosenhaus:2018zqn}
\begin{equation}
    g^{h_1,\dots,h_6}_{\mathbbm{h}_1,\mathbbm{h}_2,\mathbbm{h}_3}= 
    \chi_1^{\mathbbm{h}_1}\chi_2^{\mathbbm{h}_2}\chi_3^{\mathbbm{h}_3} 
    F_K\left[\begin{array}{c}
         h_{12}+\mathbbm{h}_1, \,\mathbbm{h}_1+\mathbbm{h}_2-h_3,\,  
         \mathbbm{h}_2+\mathbbm{h}_3-h_4  ,\, \mathbbm{h}_3+h_{65}  \\[2mm]
         2\mathbbm{h}_1,\, 2\mathbbm{h}_2,\, 2\mathbbm{h}_3
    \end{array}; \chi_1,\chi_2,\chi_3   \right]\,,
    \label{eq:one_d_six_pt_block}
\end{equation}
where the comb function $F_K$ can be expressed as
\begin{equation}
    F_K\left[\begin{array}{c}
         a_1, \,b_1,\, b_2  ,\, a_2 \\
         c_1,\, c_2,\, c_3
    \end{array}; \chi_1,\chi_2,\chi_3   \right]=
    \sum_{n=0}^\infty \frac{(b_1)_n (b_2)_n}{(c_2)_n}\frac{\chi_2^n}{n!} {}_2F_1\left[ \begin{array}{c}
         b_1+n,\, a_1  \\
          c_1
    \end{array}; \chi_1  \right] {}_2F_1\left[ \begin{array}{c}
         b_2+n,\, a_2  \\
          c_3
    \end{array}; \chi_3  \right]\,.
    \label{eq:CombFunction6pt}
\end{equation}
It is immediate to see that taking the leading behaviour on the cross ratio $\chi_2$, 
which is the one-dimensional analogue of $\bar{z}_2$, corresponds to simply 
setting $n=0$ in eq.\  \eqref{eq:CombFunction6pt}, leading to the factorised expression
\begin{equation}
    g^{h_1,\dots,h_6}_{\mathbbm{h}_1,\mathbbm{h}_2,\mathbbm{h}_3}
    \stackrel{\chi_2\rightarrow 0}{\scalebox{1.8}[1.15]{$\sim$}} 
    \chi_2^{\mathbbm{h}_2}\left(\chi_1^{\mathbbm{h}_1}{}_2F_1
    \left[ \begin{array}{c}
         \mathbbm{h}_1+\mathbbm{h}_2-h_3,\, h_{12}+\mathbbm{h}_1  \\[2mm]
           2\mathbbm{h}_1
    \end{array}; \chi_1  \right]\right) \left(\chi_3^{\mathbbm{h}_3} 
    {}_2F_1\left[ \begin{array}{c}
         \mathbbm{h}_2+\mathbbm{h}_3-h_4,\, \mathbbm{h}_3+h_{65} \\[2mm]
          2\mathbbm{h}_3
    \end{array}; \chi_3  \right]\right)\,.
    \label{eq:one_d_factorised}
\end{equation}
Expression \eqref{eq:one_d_factorised} is an explicit factorisation of six-point 
conformal blocks into the product of two four-point blocks, providing an explicit 
example of equation~\eqref{eq:bOPE_to_4pt} for the one-dimensional case.


\end{document}